%% file: paper.tex
\documentclass{sig-alternate}

\include{header}
\include{reference}

\newcommand{\DC}{\emph{DC}\xspace}
\newcommand{\bigred}{Big Red 2\xspace}
\newcommand{\emphbf}[1]{\textbf{\emph{#1}}}
\renewcommand{\footnotesize}{\small}

\DeclareGraphicsExtensions{.pdf,.png,.jpg}
\graphicspath{ {./figures/} }

\begin{document}
\conferenceinfo{Supercomputing}{2015 Austin, Texas USA}

\title{The Anatomy of Large-Scale Distributed Graph Algorithms}
\numberofauthors{5} 
\author{
\alignauthor 
Jesun Sahariar Firoz\\
       \email{jsfiroz@iu.edu}
\alignauthor
Thejaka Amila Kanewala\\
       \email{thejkane@iu.edu}
\alignauthor
Marcin Zalewski\\
       \email{zalewski@iu.edu}
\and
\alignauthor 
Martina Barnas\\
       \email{mbarnas@indiana.edu}
\alignauthor
Andrew Lumsdaine\\
       \email{lums@iu.edu}
       \and
\affaddr{Center for Research in Extreme Scale Technologies (CREST)}\\
\affaddr{Indiana University, Bloomington, IN, USA}\\
}

\maketitle

\begin{abstract}
The increasing complexity of the software/hardware stack of modern supercomputers 
has resulted in an explosion in the parameter space for performance tuning.
In many ways performance analysis has become an experimental science, made even more challenging due to the presence of massive irregularity and data dependency in important emerging problem areas.
As with any experimental science, a characterization of experimental conditions is important for identifying which variables in the experiment affect the outcome.
To gain insight into the experimental nature of performance analysis, we analyze how the existing body of research handles the particular case of distributed graph algorithms (DGAs).
We distinguish \emph{algorithm-level} contributions, often prioritized by authors, from \emph{runtime-level} concerns that are harder to place. 
With a careful exposition of the tuning process for a high-performance graph algorithm, 
we show that the runtime is such an integral part of DGAs that experimental results are difficult to interpret and extrapolate without understanding the properties of the runtime used. We argue that in order to gain understanding about the impact of runtimes, more information needs to be gathered. To begin this process, we provide an initial set of recommendations for describing DGA results based on our analysis of the current state of the field.
\end{abstract}





\newpage

\section{Introduction}
\label{sec:intro}
Large irregular applications are gaining recognition as the future challenge in parallel computing. This is reflected by the Graph500 benchmark~\citep{murphy2010introducing}, the subject of which is the prototypical irregular problem of graph traversal.  Graph traversal is a basic building blocks of other graph algorithms used in social network analytics, transportation optimization, artificial intelligence, power grids,  and, in general, any problem where data consists of entities that connect and interact in irregular ways.  The current Graph500 benchmark is  based on breadth-first search (BFS) with a proposal to extend the benchmark with single-source shortest paths (SSSP). 
In this paper, we concentrate on BFS and SSSP for the same reasons, i.e., as representatives of a class of irregular graph problems.

Research on distributed graph algorithms is an emerging and active field. New algorithms, new approaches to distribute the data and new performance results appear at most major distributed computing conferences. The Graph500 benchmark bears witness to the progress, with the best results progressing from 7 GTEPS (billions of traversed edges per second) in 2010, 253 GTEPS in 2011, 15 TTEPS (trillions of TEPS) in 2012, to 23 TTEPS in 2014. Many new algorithmic techniques have been developed, \eg direction optimization~\citep{beamer_direction-optimizing_2013,beamer_distributed_2013}, pruning \citep{chakaravarthy_scalable_2014}, k-level asynchronous algorithm \citep{harshvardhan_kla:_2014}, hybrid algorithms \citep{chakaravarthy_scalable_2014}, and distributed control \citep{zalewski_distributed_2014}.  A practitioner faces a multitude of published approaches, which are often vague on low-level details of implementations.

Graph problems are difficult to fit to the conventional High Performance Computing (HPC) platforms. Over the decades of development, these platforms have been optimized for problems exhibiting good locality and regular memory access and communication patterns, benefiting from caching and high-bandwidth regular collective operations. In contrast, graph algorithms exhibit little locality, rarely require any significant computation per memory access, and result in high-rate communication of small messages. 
Unlike regular applications that are built on top of well understood regular communication and memory access, graph algorithms interact with the whole software and hardware stack in a complex way due to their data-driven, fine-grained, irregular nature of tasks.  
Each piece of the stack, designed independently, from the application level, through the transport layer, to the hardware layer and the topology of the 
physical network, interacts within the system in unpredictable ways.  This makes designing distributed graph algorithms a truly experimental science, and this state of affairs will be only exacerbated as we move towards exascale computing.


We argue that the advancements in the field are hard to generalize and reconcile because the information reported is commonly incomplete. The low-level details of implementations are often vague or missing.  Yet, these can have important impact. 
In this paper we present evidence of impact of low-level transport, scheduler, and hardware, which we refer to as  \emph{runtime}. It should be noted that the complexity of the interactions between high-level algorithm and low-level runtime that we expose is not unknown to the scientists in the field.  This knowledge is implicit, fragmented, and often sidelined in presentation of new techniques.  Notably, \citet{checconi_traversing_2014}, who achieve the top results in the Graph 500 benchmark in part due to direct access to the SPI (System Programming Interface) low-level primitives, provide an outstanding analysis of their evolving implementation, including a 3-years timeline of changing conclusions and understanding. However, this is not typical for the field.  We point out a need for community standards and guidelines for presenting experimental results.



\vspace{1ex}
\noindent\textbf{Contributions:}

\textbf{A motivating case study} (\secref{sec:motivation});

\textbf{A survey and analysis of the field} (\secref{sec:anatomy}), identifying, classifying, and discussing two levels of distributed graph algorithms: (i) Application-level aspects (\cref{sec:application}) that authors identify as the main algorithmic contributions of their research; and (ii) Runtime-level aspects (\cref{sec:runtime}) that authors do not explicitly consider a part of the algorithm but that play a crucial role in the overall performance;

\textbf{An initial set of recommendations} (\cref{sec:recommendations}) for authors to consider when describing their research.

\section{Motivating Case Study}
\label{sec:motivation}
\citet{pingali2011tao} classify algorithms into two main categories: \textit{ordered} and \textit{unordered}.  Ordered algorithms require ordering of tasks for correctness whereas unordered algorithms do not.  Parallelizing ordered algorithms is challenging as parallel execution must maintain the ordering. Unnordered algorithms are easier to parallelize as tasks can be executed in any order.  Although the order of execution does not impact correctness in unordered algorithms, a \emph{task priority} can be used to partially order tasks, improving performance.  Our \textit{distributed control} (\DC) approach \citep{zalewski_distributed_2014} is a work scheduling method for distributed unordered algorithms that benefit from task priority.  The goal of \DC is to remove the overhead of synchronization and global data structures by using only local knowledge to select best work, thus obtaining an approximation of the global ordering.  Because \DC does not use global data structures and synchronization, it is particularly sensitive to the runtime characteristics such as timing of task execution, communication latency, and buffering.   In fact, their effect can be so significant as to make \DC an infeasible approach if not handled carefully.  In this section, we discuss 
the significant runtime effects we have observed.

\begin{figure}
\centering
\includegraphics[width=\linewidth]{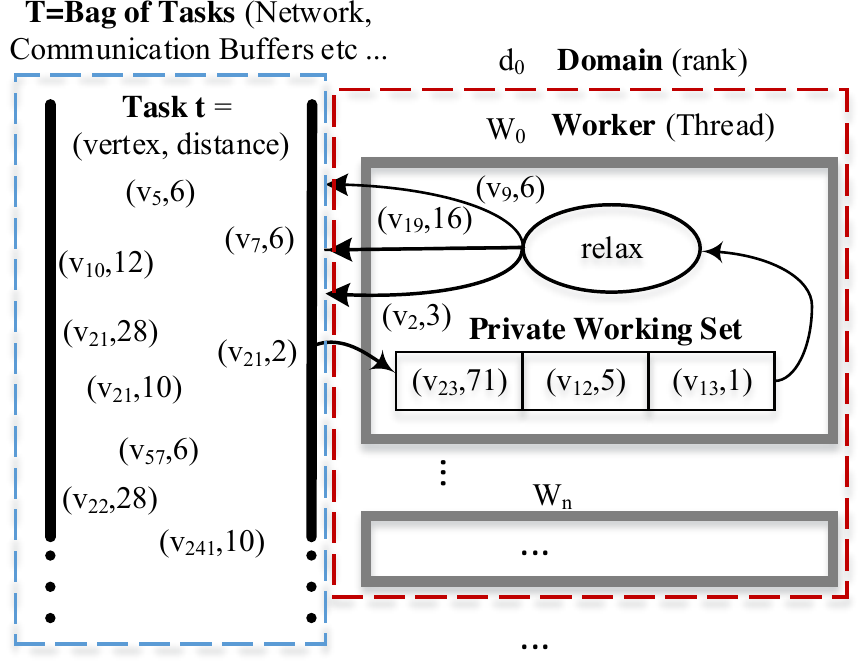}
\caption{An overview of \textit{Distributed Control}, using SSSP as an example.}
\vspace{-1ex}
\label{fig:dc}
\end{figure}

\Cref{fig:dc} illustrates \DC for SSSP.  A distributed system consists of workers divided into several shared memory domains.  Each domain contains a part of the global data.
Processing a task on one domain may generate more tasks that depend on data on other domains. Remote tasks are communicated through an unordered \textit{global task bag} where every worker puts the tasks it generates.  Workers continuously try to retrieve tasks from the bag into their \textit{private working sets} that are ordered according to task priority.  Because the task bag is unordered, the more tasks in the bag and not in the ordered private priority queues of workers, the further the approximated ordering is from the ideal least-work ordering (Dijkstra's priority queue).  

Ideally, the underlying runtime system delivers tasks to the appropriate ordered private workset as soon as possible.  On the other hand, quick delivery comes at a cost: the accumulative costs of network sends overhead, the necessity for frequent polling, frequent context switches when handling small tasks, and so on, add up to a significant overhead.  To balance these competing needs, we use the \ampp~\citep{willcock_am++:_2010} runtime. \ampp is particularly suitable for this purpose because it supports fine-grained parallelism of active messages with communication optimization techniques such as scalable addressing, active routing, message coalescing, message reduction, and termination detection.  

In the remainder of this section, we discuss the characteristics of the runtime we experimented with and the impacts we observed.  Our experiments were run at different times on different machines (Cray environments).  The experiments we discuss here were ran on \bigred at Indiana University~\citep{bigred2} and on Edison at NERSC~\citep{nerscedision}.  The most important differences between the two machines are the topologies, 3-D torus on \bigred vs. Dragonfly on Edison, and the MPI implementations, Cray's Message Passing Toolkit (MPT) 6.2.2 on \bigred vs. MPT 7.1.1 on Edison.  All experiments were run on Graph500 graphs.

\subsection{Runtime Parameters}
\label{sec:runtime-params}
\DC consists of layers, each with its own set of possible design choices and parameters.  The exact set of parameters suitable for an application depends on the specifics of the machine (network overheads, etc.) and on the input (graph structure, edge weights, etc.).  Some examples of parameters in our implementation include:
\begin{itemize}
\item Progress related parameters such as the threshold for eager progress and the frequency of progress calls.
\item Optimizations such as reduction cache size, priority messaging, and ``self-send'' check.
\item Coalescing buffers size, controlling the maximum number of messages that can be sent together.
\item MPI-related parameters such as receive depth, the number of polling tasks, and the number of outstanding sends allowed.
\end{itemize}
We discuss some of these parameters in more detail in the remainder of this section.  In addition to parameters directly related to \ampp and our algorithm, there are environment parameters, including:
\begin{itemize}
\item MPI progress model, with several choices on the machines that we have run on.
\item Job placement, which can severely impact performance.
\item Use of huge pages modules on Cray machines.
\end{itemize}
The parameter space is further enlarged by the characteristics of inputs. For example, we observed that edge weights have a significant impact on the choice of optimal parameters for \DC.

\subsection{Coalescing Size}\label{section_coalescing_size}
\label{sec:coalescing-size}
\setlength{\intextsep}{1.5ex}%
\setlength{\columnsep}{1.5em}%
\begin{figure}
\centering
\includegraphics[width=.6\linewidth]{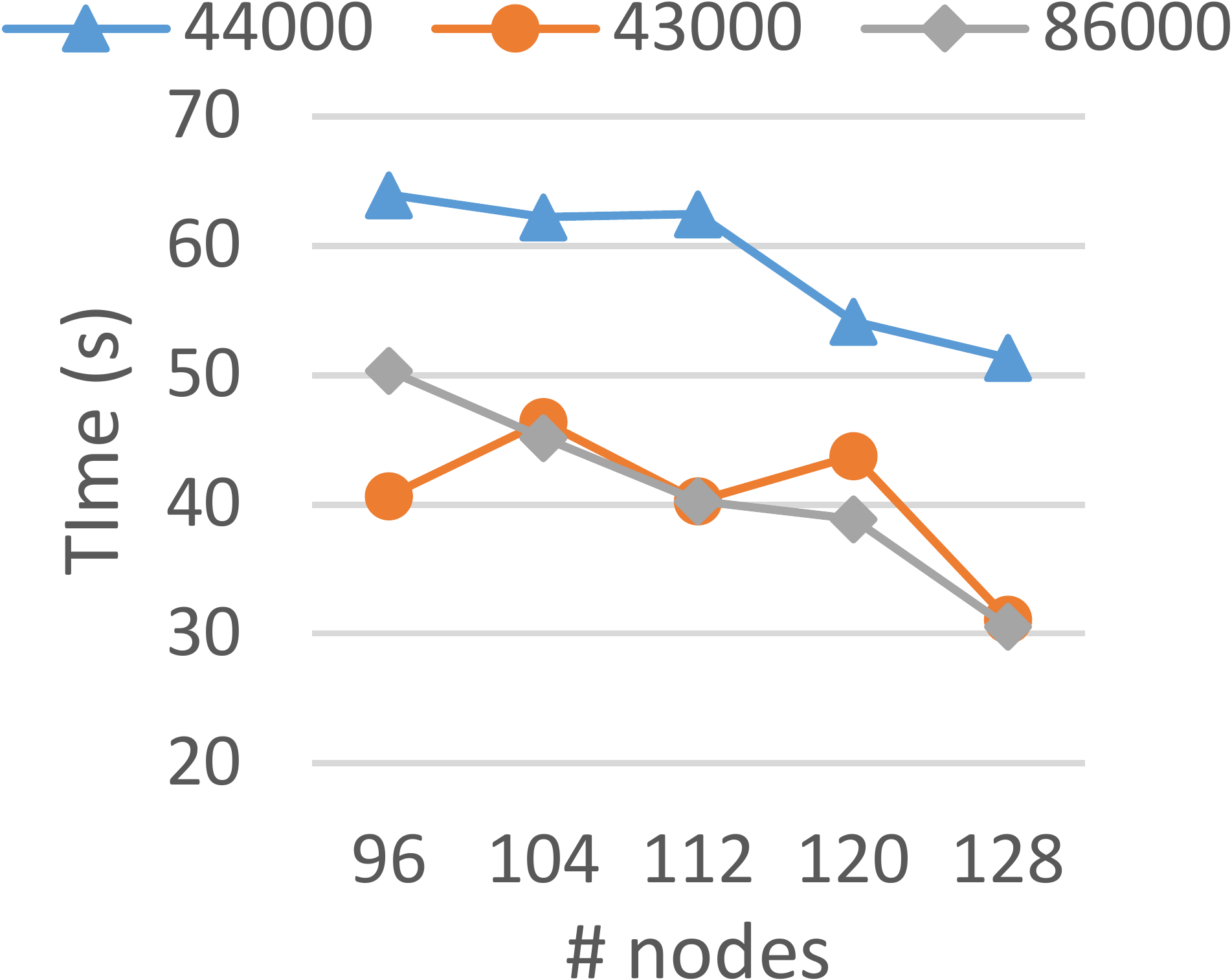}
\caption{Effect of coalescing size on \DC SSSP algorithm on a scale 31 graph (\bigred).}
\label{fig:coalescing}
\end{figure}
To increase bandwidth utilization, \ampp performs \textit{message coalescing}, combining multiple messages sent to the same destination into a single, larger message.  Messages are appended to per-destination buffers, and to handle partially filled buffers, a periodic check is performed to check for activity.  In the case of \DC SSSP, a single message consists of a tuple of a destination vertex and distance, 12 bytes in total.  With such small messages, coalescing has great impact on the performance, but finding the optimal size is difficult.

\begin{figure}
\centering
\includegraphics[width=\linewidth]{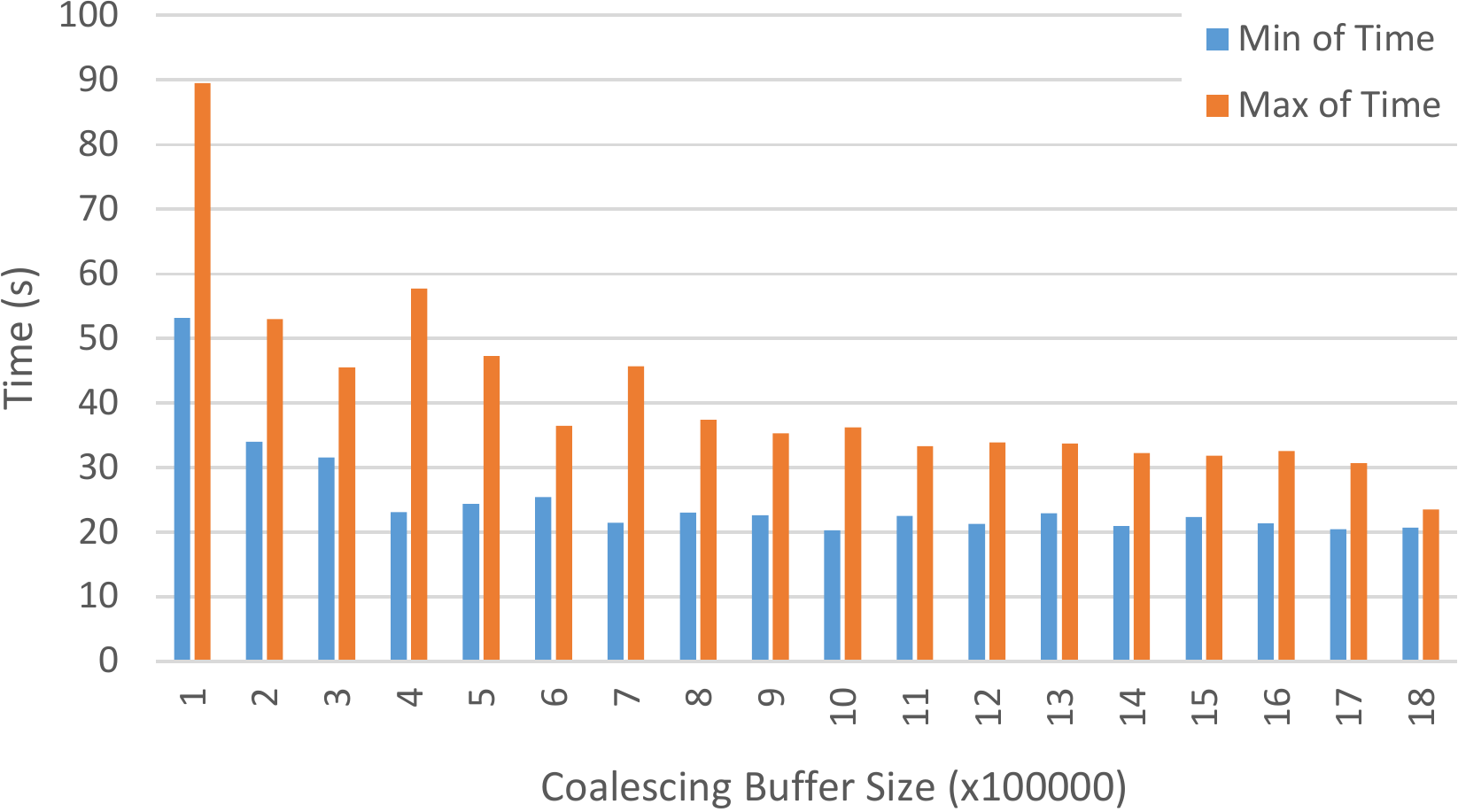}
\caption{Effect of coalescing size on \DC SSSP algorithm on a scale 31 graph (Edison).}
\label{fig:edison-coalescing}
\end{figure}

\begin{figure}
\centering
\includegraphics[width=.5\linewidth]{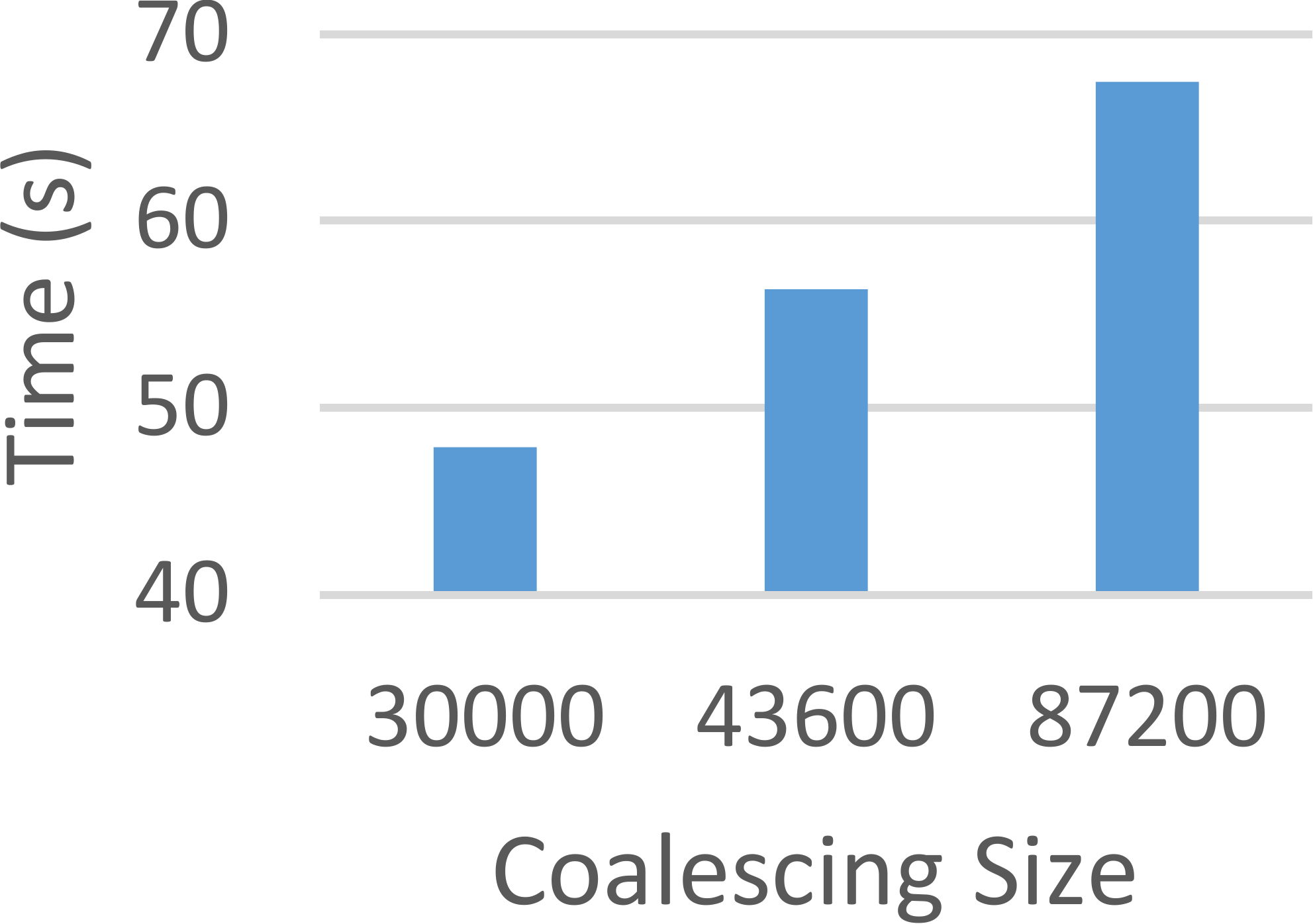}
\caption{Effect of coalescing size on \DC BFS algorithm on a scale 31 graph (\bigred).}
\label{fig:bfs-coalescing}
\end{figure}

We investigated the impact of coalescing in graph500 scale 31 graphs when running \DC SSSP with max edge weight of 100(\cref{fig:coalescing,fig:edison-coalescing}).  \Cref{fig:coalescing} shows the large impact of a small change in the coalescing size, measured by the number of SSSP messages per coalescing buffer. Changing the coalescing size by less than 2\% causes over 50\% increase in the run time.  This unexpected effect is caused by the specifics of Cray MPI protocols.  At the smaller coalescing size, full message buffers fit into rendezvous R0 protocol that sends messages of up to 512K using one RDMA GET, while the larger buffers hit R1 protocol that sends chunks of 512K using RDMA PUT operations.  At the size of 44000, the bulk of the message fits into the first 512K buffer, and the small remainder requires another RDMA PUT, causing overheads.  The sizes 43000 and 86000 fill out 1 and 2 buffers, respectively, achieving similar performance.  The larger size, 86000, results in better scaling properties.

We ran a more extensive suite of benchmarks on Edison.  \Cref{fig:edison-coalescing} shows the coalescing buffer size experiments on Edison. The results are similar, with a periodic increases in the minimum run time as protocol buffers mismatch the coalescing buffers.  The maximum run times signify the worst run time, as other parameters than coalescing are adjusted.  The results show that adjusting other parameters is less and less important as the coalescing buffer size increases.

\label{sec:transport-progress}
\begin{figure}
\centering
\includegraphics[width=.6\linewidth]{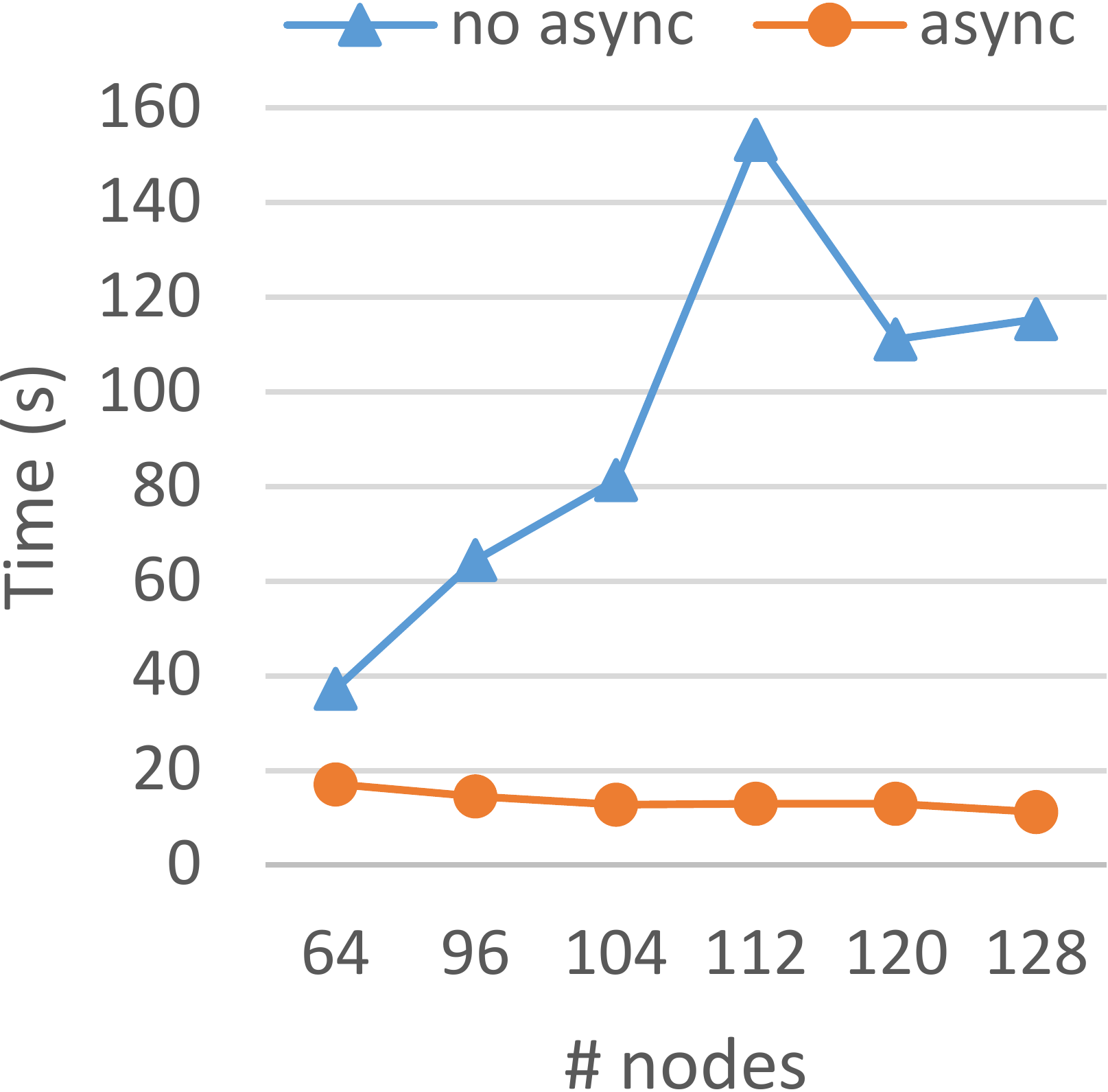}
\caption{Effect of asynchronous progress on \bigred.}
\label{fig:async}
\end{figure}

\Cref{fig:bfs-coalescing} shows the effects of coalescing on a \DC BFS, which is SSSP with maximum weight of 1.  Surprisingly, increasing the coalescing size impacts performance negatively.  We suspect that with smaller weights the possibility of reward from optimistic parallelism in \DC decreases, and the added latency of coalescing has a much larger effect than with larger weights.  Also note that we have not actually discovered the optimal coalescing size, which would require more experiments and more resources.  All three cases shown in \cref{fig:coalescing,fig:edison-coalescing,fig:bfs-coalescing} show that adjusting the coalescing size is important, and the optimal value is not static. Rather, it depends on algorithmic concerns such as reward from optimistic parallelism.

\subsection{Transport Progress}\label{sec:dc-progress}
At first, when we experimented with \DC on \bigred, we found out that it was performing worse than \Dstepping. This raised a concern that the \DC approach may not be practical.  We suspected the possibility of message latencies being a culprit; so, upon researching MPT, we decided to experiment with asynchronous progress, which uses separate threads to perform progress in certain situations.  Despite Cray's warning at the time that thread-multiple progress required for asynchronous progress ``is not considered a high-performance implementation'', we observed significant gains for \DC, shown in \cref{fig:async}.  We ran the experiment on Graph500 scale 31 optimal strong scaling results.  Without asynchronous progress, performance decreased with the increased number of nodes (with an unexplained anomaly at 112 nodes). (Note that all our experiments are averaged; thus, large anomalies are indicative of unexpected circumstances.)  With asynchronous progress thread, the performance of \DC has improved more than tenfold with growing node counts, entirely changing the viability of the approach.  This dramatic effect illustrates how deeply an algorithm interacts with the runtime, and how a gap in parameter space may lead to incorrect conclusions about DGA approaches.  Interestingly, we did not observe a similar effect on Edison, where two different asynchronous progress and the standard progress modes perform similarly.

\subsection{Distributed Control Progress}
\label{sec:dc-progress}
\begin{figure}
  \centering
  \begin{subfigure}[b]{.4\linewidth}
    \centering
    \includegraphics[width=.89\linewidth]{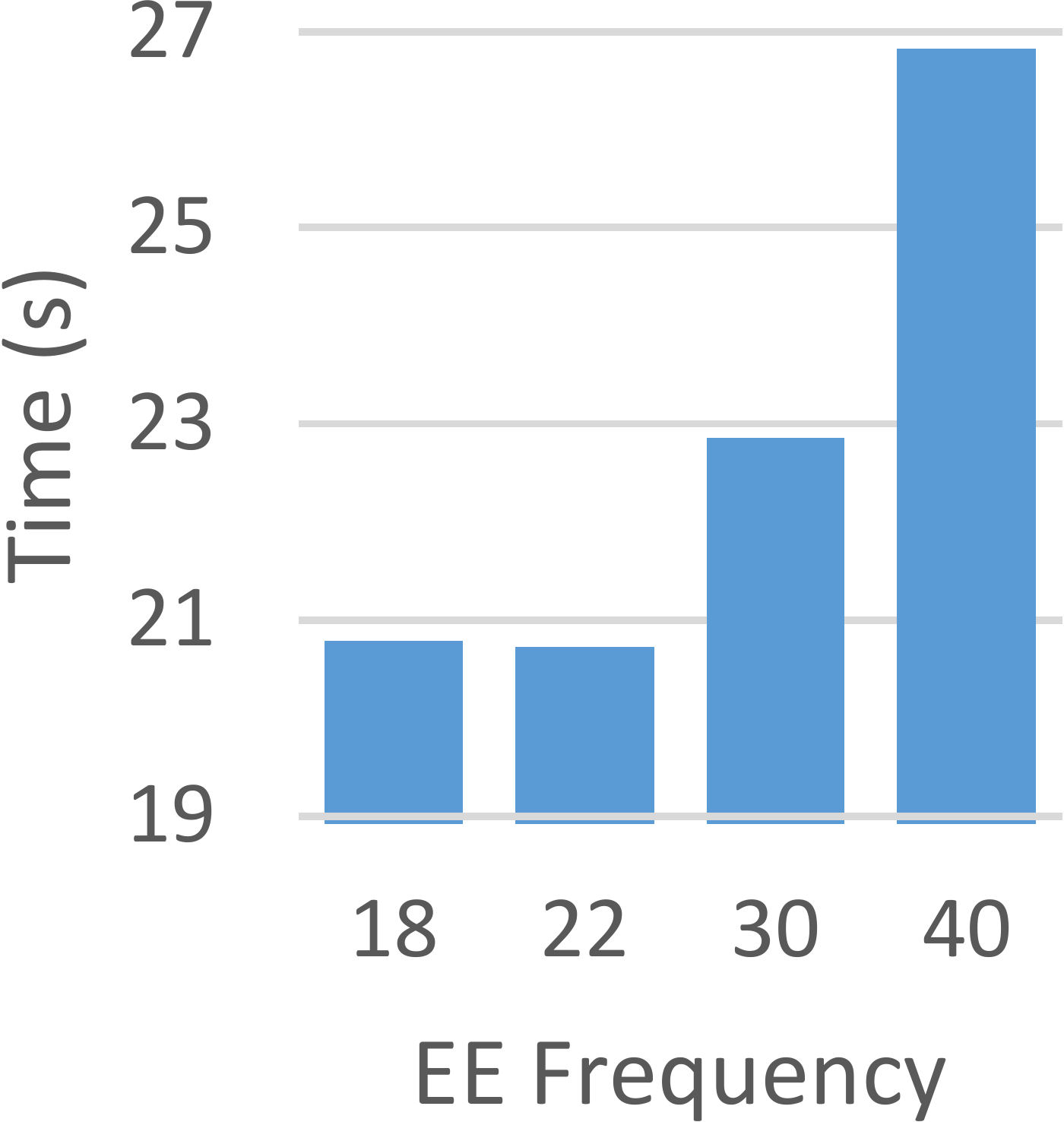}
    \caption{EE on Edison.}
  \end{subfigure}\hspace{2em}
  \begin{subfigure}[b]{.4\linewidth}
    \centering
    \hspace{-.89em}
    \includegraphics[width=\linewidth]{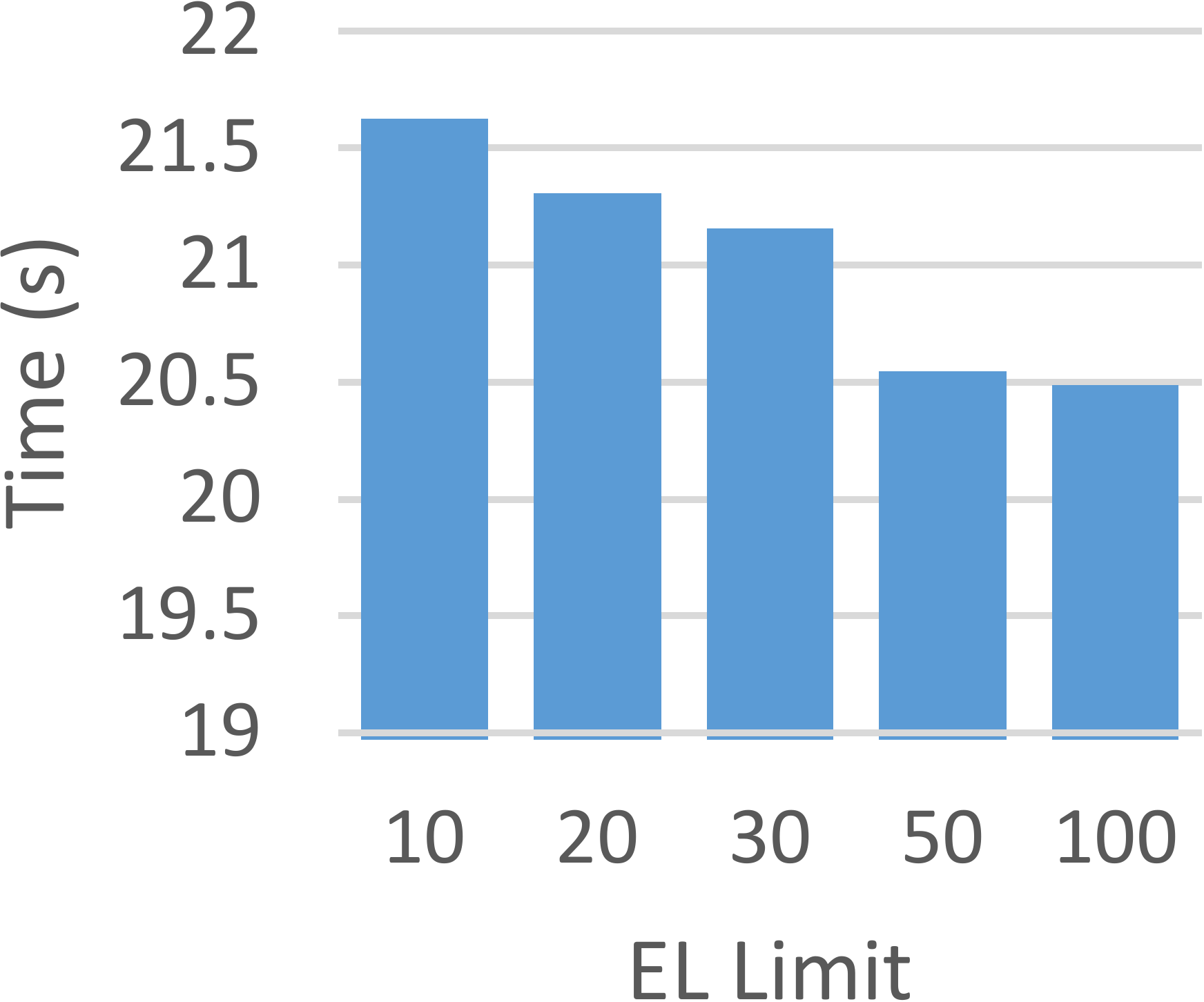}
    \caption{EL on Edison.}
  \end{subfigure}\\[2ex]
  \begin{subfigure}[b]{.4\linewidth}
    \centering
    \includegraphics[width=\linewidth]{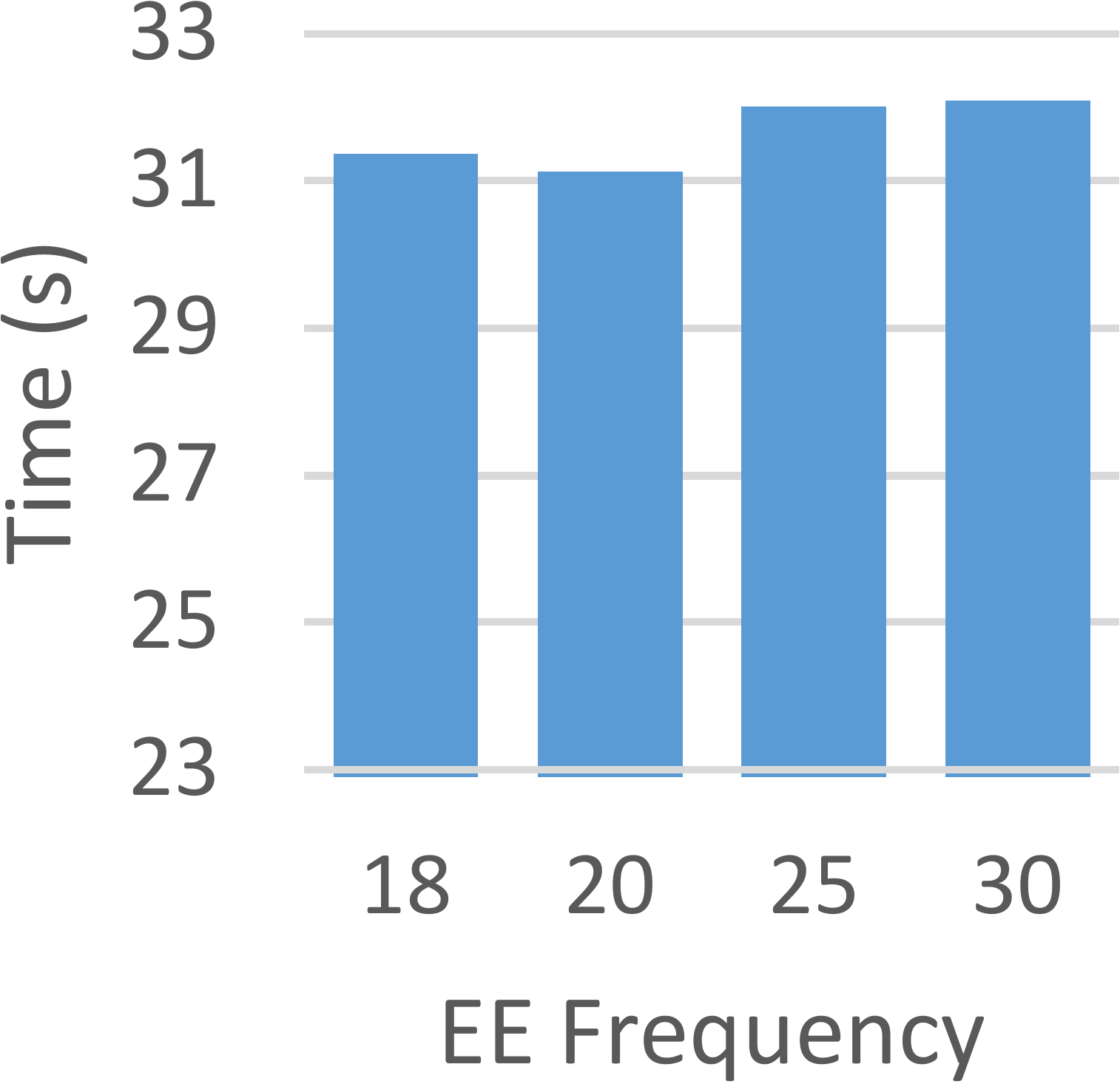}
    \caption{EE on BR2.}
  \end{subfigure}
  \caption{Effect of AM++ progress parameters.}
  \label{fig:ampp-progress}
\end{figure}
In addition to transport layer progress, \ampp performs its own internal progress when \ampp interfaces are called.  Since \DC is built around a loop that empties the local priority data structure, it must occasionally, with some frequency call into the appropriate \ampp interfaces that perform progress.  This frequency is controlled by 2 parameters: the end-epoch test frequency (EE) and the eager progress limit (EL).  EE controls how many iterations of the \DC loop run before \ampp progress is invoked. The eager limit is a threshold of outstanding \DC tasks below which \ampp progress is performed every iteration of the \DC loop.

\Cref{fig:ampp-progress} shows the effects of progress parameters, using performance data averaged over multiple runs while varying orthogonal parameters.  Edison shows a significant sensitivity to the EE parameter.  Smaller values are better, with 22 being the best of the ones tested.  This suggests that latency may be a limiting factor on Edison.  On \bigred, the results of varying the EE parameter are less pronounced, but the average of multiple experiments that we show here still suggests some sensitivity with the optimal value similar to that on Edison.  Altogether, the results show that the performance of \DC depends on the progress model.

\subsection{Buffering and Work Efficiency}\label{section_partial_and_full_buffers}
\label{sec:buffering}
\MZ{Maybe talk about: * smaller depth works better and benefits from increased polling. larger depths exhibit the opposite effect.}
\begin{figure}
  \centering
  \includegraphics[width=\linewidth]{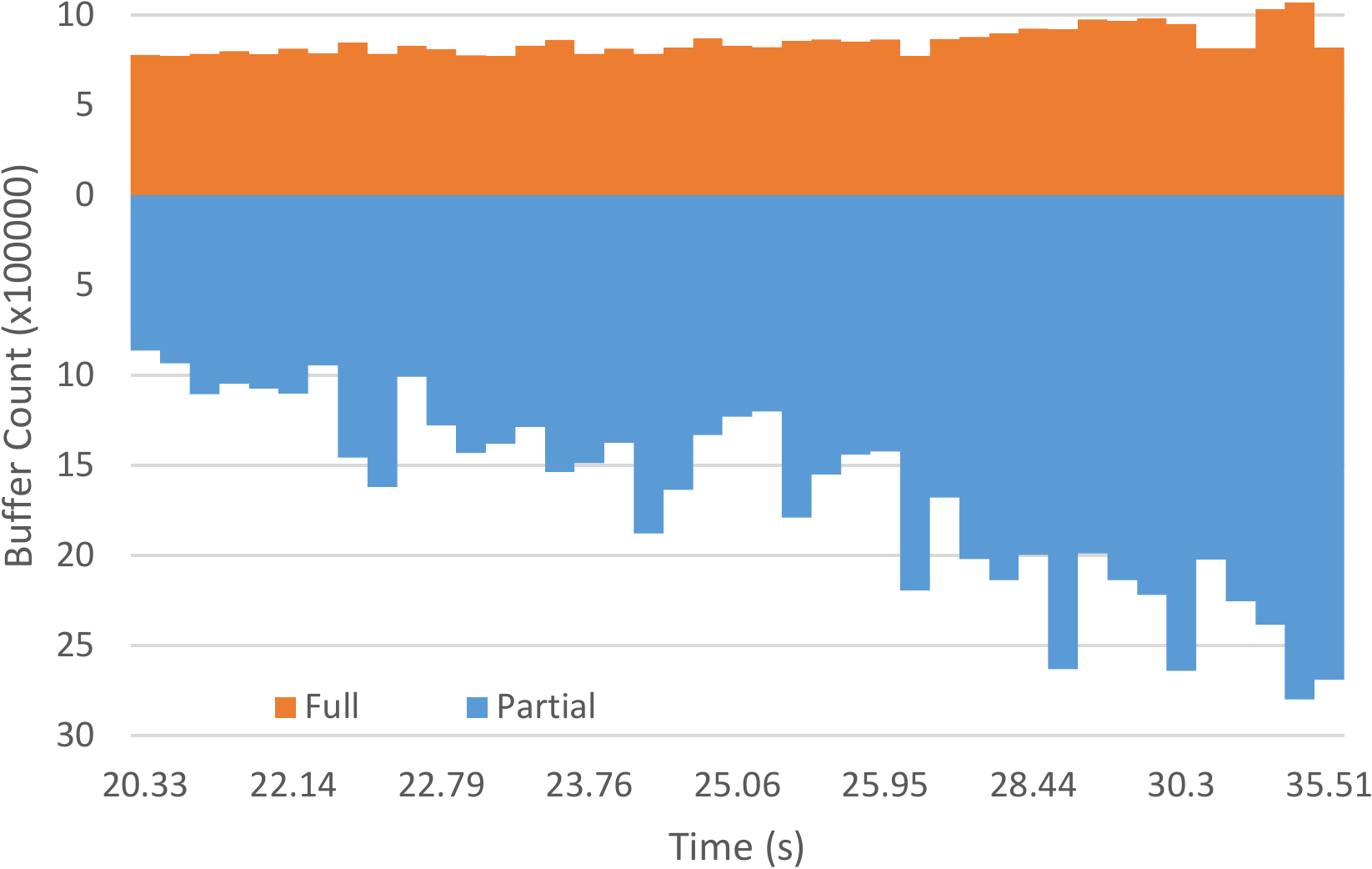}
  \caption{The impact of partial and full buffer counts on performance with coalescing size fixed at 100000 (Edison).}
  \label{fig:buffers}
\end{figure}

The prerogative of coalescing in \ampp is to decrease the overhead by sending as many full coalescing buffers as possible.  Partially filled buffers are only sent when no more messages are being inserted.  \Cref{fig:buffers} shows \DC results on Edison for coalescing buffer size of 100000.  We found that the best predictor of performance is the amount of partial buffers (fewer is better) followed by full buffers (more is better).  Partial buffers indicate periods of lack of work, and this, in turn, indicates that the local priority queues are getting depleted more often, decreasing overall performance.  \ampp was originally optimized for algorithms like BFS and \Dstepping, which benefit from eager optimization of communication overhead and are not sensitive to work imbalance.  Our example shows that optimization of runtime for a seemingly worthy goal can negatively impact applications that have other needs not anticipated by runtime developers.

\subsection{Performance Irregularity}

\begin{figure}
  \centering
  \includegraphics[width=.6\linewidth]{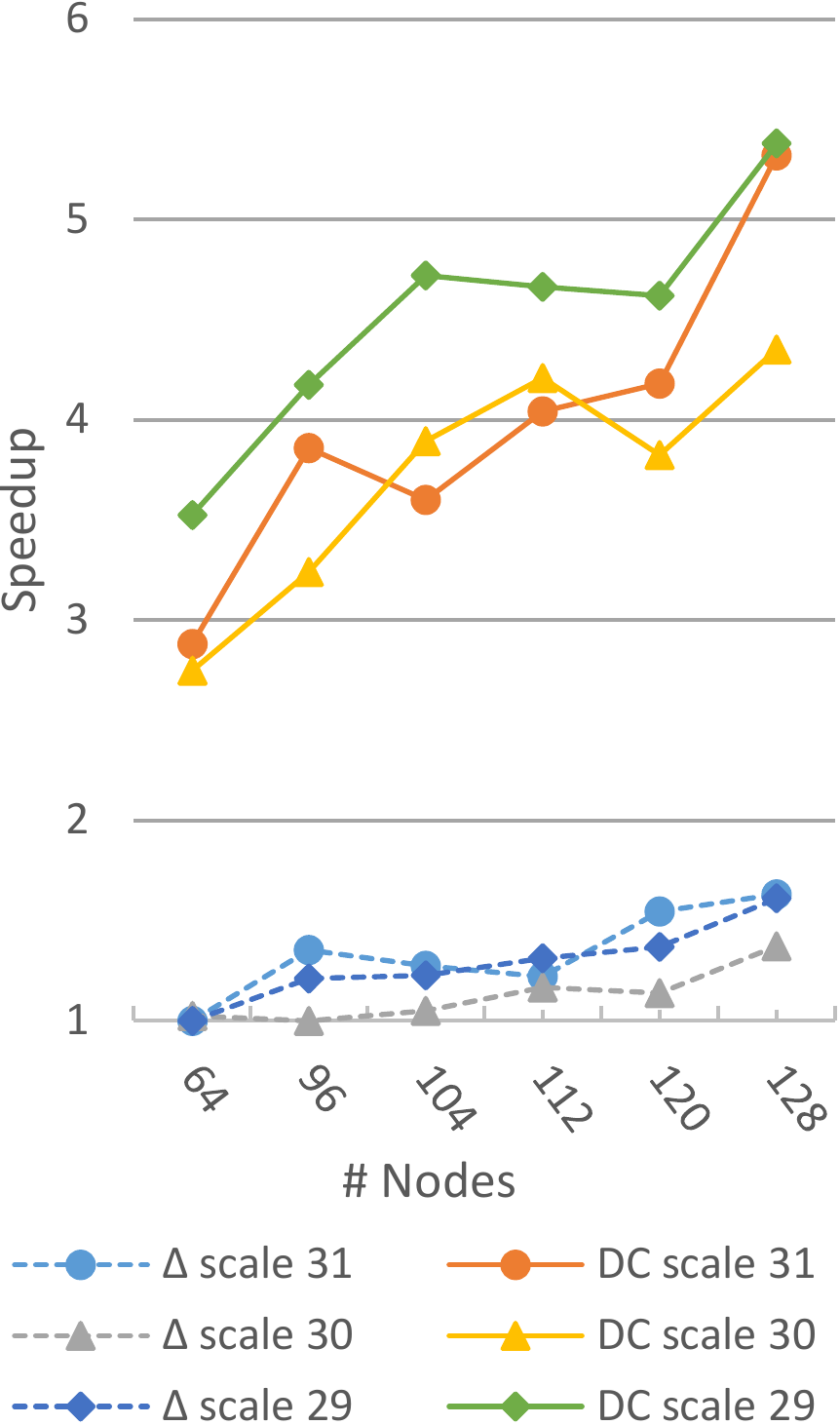}
  \caption{Irregularities in strong scaling on \bigred.}
  \label{fig:scaling}
\end{figure}

A large supercomputer runs jobs on a complex topology, with an allocation system designed to maximize utilization and ensure fairness.  To satisfy these goals and to account for complex topologies, jobs may be mapped to hardware in different ways between different runs, and hardware resources may be shared by jobs in different degrees depending on the current job configuration.  
\Cref{fig:scaling} shows scaling results for \DC together with our implementation of \Dstepping for comparison.
\DC experiences a drop in performance after 96 nodes, as does \Dstepping, although \DC recovers faster than \Dstepping.  A similar drop can be observed for scale 30, but it occurs later, at 112 nodes.  \DC for scale 29 also shows a drop in performance after 104 nodes, but it is not as dramatic as at larger scales.  Currently, we do not have an explanation for performance variability, but we suspect the effects of job placement as discussed in \cref{sec:network-topo}, where at some node counts a job on a certain input does not map well to the underlying topology.

Another kind of irregularities are not as consistent, and they depend on transient conditions such as availability of contiguous blocks of nodes and interaction with other jobs.  For example, on Edison, we noticed that some jobs we run more than once took up to 25\% more time in some runs.

\subsection{Caching}\label{sorting_and_caching}
We implemented a write-through cache with the most recent SSSP messages.  When a message to a given destination is discovered in cache, it is either discarded if the previous message had smaller distance, or it replaces the old message in cache and is sent through.  Caching can improve performance dramatically, as we observed in our implementation of \Dstepping.  However, for \DC, with caching performance deteriorates.  For example, on runs on scale 30 graph on \bigred, \DC performed by almost 30\% worse with caching despite reducing work by about 10\%.  The interesting conclusion is that an optimization technique can be actually detrimental to the performance of an algorithm, which can be counterintuitive.

\subsection{Work vs. Overhead}

\begin{figure}
\centering
\includegraphics[width=\linewidth]{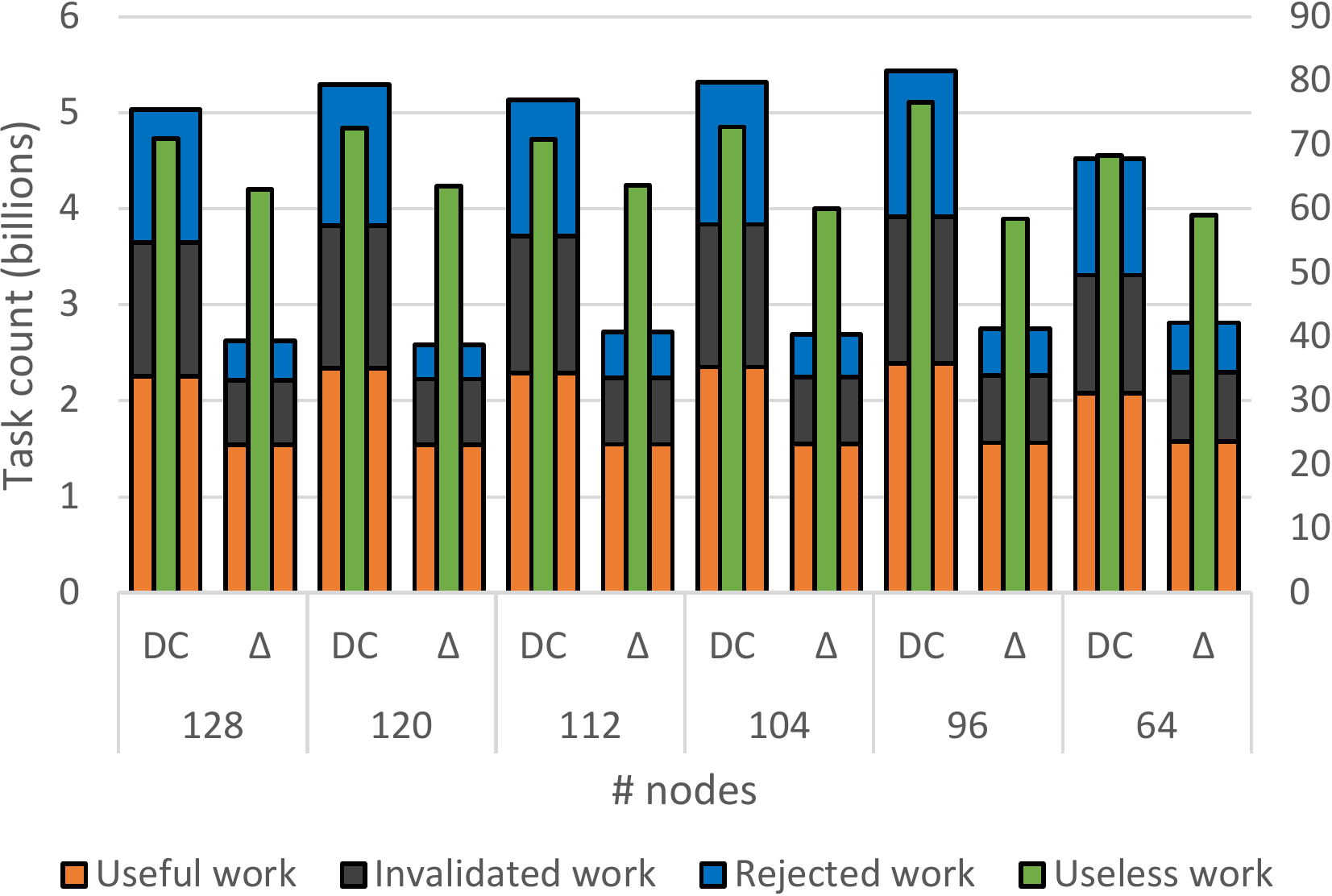}
\caption{Proportion of different kind of works in DC-SSSP and $\Delta$-stepping algorithm.}
\label{fig:work-stat}
\end{figure}%

Performance of an algorithm depends on the amount of \emph{work} it performs and on the amount of \emph{overhead} that this work incurs in a given runtime. \Cref{fig:work-stat} shows the work statistics comprising of useful work, useless work, rejected work and invalidated work for DC and our implementation of \Dstepping with scale 31 graph. While DC performs better than \Dstepping, DC always executs more work than \Dstepping in the most efficient configurations of both of the algorithms. As can be seen from \cref{fig:work-stat,fig:scaling}, despite consistently performing 10\%-25\% more work, DC performs better in all instances of tests at scale.  This shows that synchronization and uneven distribution of work have an important effect on the performance of DGAs.  While one can attempt to mitigate the work imbalance with algorithmic techniques, the cost of synchronization is hard to control and eliminate. In this regard, an underlying runtime can have a significant impact. The more an algorithm depends on keeping global information about the runtime (\eg for load balancing), the higher the costs of synchronization necessary to maintain that information. In \Cref{fig:work-stat} we count a task as rejected when the vertex distance in delivers is higher than what is already recorded and, consequently, the task is not inserted into the priority queue of DC or a bucket of \Dstepping.  Invalidated tasks are similar to rejected tasks, but their distance expires while they wait in priority queue.

\section{The Anatomy of DGA\lowercase{s}}
\label{sec:anatomy}
In previous section we show the interactions between runtime and the application cannot be neglected.  Next we analyze the existing research on distributed BFS and SSSP problems.  We provide \emphbf{the anatomy} of DGAs, consisting of \emph{application-level} (\cref{sec:application}) and \emph{runtime-level} (\cref{sec:runtime}) aspects.  As explained in \cref{sec:intro}, this division is based on how the research results are presented in the field.  Authors usually concentrate on the application-level aspects, framing their contributions at that level, and treat the interactions with runtime as secondary results, often providing incomplete information about it.  The purpose of our analysis is to describe the complexity of interactions between application and runtime and to provide a blueprint for a more complete treatment of DGAs.

\subsection{Application-Level aspects of DGAs}
\label{sec:application}
\begin{table}
  \centering
  \small
  \setlength{\tabcolsep}{8pt}
  \begin{tabular}{@{}p{.3\linewidth}@{}p{.9\linewidth}@{}}
    \toprule
    \addlinespace
    \textbf{Approach} & \multicolumn{1}{@{}r@{}}{\cref{sec:approach}}\\
    \midrule
    \LGlcell{} & \LGrcell{Level-Synchronous\textsuperscript{\citep{yoo_scalable_2005, checconi_traversing_2014}}} \\
    \LGlcell{Ordered} & \LGrcell{Bellman-Ford\textsuperscript{\citep{madduri_et_al_performance_2014}}}  \\
    \LGlcell{} & \LGrcell{Combinatorial BLAS\textsuperscript{\citep{buluc_parallel_2011, beamer_direction-optimizing_2013, buluc_distributed-memory_2015}}}  \\
    \MGlcell{} & \MGrcell{Hybrid\textsuperscript{\citep{chakaravarthy_scalable_2014}}}  \\
    \MGlcell{\multirow{-2}{*}{Ordered/unordered}} & \MGrcell{HSync\textsuperscript{\citep{xie_sync_2015}}}  \\
    \LGlcell{} & \LGrcell{Distributed control\textsuperscript{\citep{zalewski_distributed_2014}}}  \\
    \LGlcell{Unordered} & \LGrcell{KLA\textsuperscript{\citep{harshvardhan_kla:_2014}}}  \\
    \LGlcell{} & \LGrcell{\Dstepping\textsuperscript{\citep{edmonds_single-source_2006, madduri_et_al_performance_2014}}}  \\

    \addlinespace[2\defaultaddspace]
    \mbox{\textbf{Algorithmic Considerations}} & \multicolumn{1}{@{}r@{}}{\cref{sec:algorithmic}} \\
    \midrule
    \LGlcell{} & \LGrcell{1D\textsuperscript{\citep{edmonds_single-source_2006, madduri_et_al_performance_2014, chakaravarthy_scalable_2014, zalewski_distributed_2014, checconi_traversing_2014}}}  \\
    \LGlcell{Data Distribution} & \LGrcell{2D\textsuperscript{\citep{yoo_scalable_2005, buluc_parallel_2011, buluc_distributed-memory_2015}}}  \\
    \LGlcell{} & \LGrcell{Edge list\textsuperscript{\citep{pearce_scaling_2013, pearce_faster_2014}}}  \\
    \MGlcell{} & \MGrcell{Ghosts\textsuperscript{\citep{gonzalez2012powergraph, low2012distributed, pearce_scaling_2013, pearce_faster_2014}}}  \\ 
    \MGlcell{} & \MGrcell{Direction optimization\textsuperscript{\citep{chakaravarthy_scalable_2014, beamer_direction-optimizing_2013, checconi_traversing_2014, buluc_distributed-memory_2015}}}  \\ 
    \MGlcell{} & \MGrcell{Pruning\textsuperscript{\citep{chakaravarthy_scalable_2014}}}  \\ 
    \MGlcell{\multirow{-2}{*}{Optimizations}} & \MGrcell{Priority messages\textsuperscript{\citep{zalewski_distributed_2014}}}  \\ 
    \MGlcell{} & \MGrcell{Tree-based broadcast, reduction,} \\
    \MGlcell{} & \MGrcell{\hspace{2em}and filtering\textsuperscript{\citep{pearce_faster_2014}}} \\
    \LGlcell{} & \LGrcell{Per-thread work splitting\textsuperscript{\citep{chakaravarthy_scalable_2014}}}  \\
    \LGlcell{} & \LGrcell{Random shuffling of} \\
    \LGlcell{} & \LGrcell{\hspace{2em}vertex identifiers\textsuperscript{\citep{buluc_parallel_2011}}} \\
    \LGlcell{\multirow{-2}{*}{Load Balancing}} & \LGrcell{Delegates\textsuperscript{\citep{pearce_faster_2014}}}  \\
    \LGlcell{} & \LGrcell{Proxies\textsuperscript{\citep{chakaravarthy_scalable_2014}}}  \\
    

    \addlinespace[2\defaultaddspace]
    \mbox{\textbf{Graph Representation}} & \multicolumn{1}{@{}r@{}}{\cref{sec:graph_representation}} \\
    \midrule
    \multicolumn{2}{@{}>{\columncolor{LightGray}[0pt][\tabcolsep]}l}{\mbox{CSR\textsuperscript{\citep{buluc_parallel_2011, beamer_direction-optimizing_2013, madduri_et_al_performance_2014, zalewski_distributed_2014, buluc_distributed-memory_2015}}, compressed coarse-index adjacency list\textsuperscript{\citep{checconi_traversing_2014}},}}  \\
    \multicolumn{2}{@{}>{\columncolor{LightGray}[0pt][\tabcolsep]}l}{\mbox{skip list\textsuperscript{\citep{checconi_traversing_2014}}, doubly-compressed sparse column (DCSC)\textsuperscript{\citep{buluc_parallel_2011,buluc_distributed-memory_2015}}}}  \\

    \addlinespace[2\defaultaddspace]
    \mbox{\textbf{Data Structures (algorithm progress)}} & \multicolumn{1}{@{}r@{}}{\cref{sec:datastructures}} \\
    \midrule
    \multicolumn{2}{@{}>{\columncolor{LightGray}[0pt][\tabcolsep]}l}{\mbox{Distributed async visitor queue\textsuperscript{\citep{pearce_scaling_2013, pearce_faster_2014}}, thread-local}}  \\
    \multicolumn{2}{@{}>{\columncolor{LightGray}[0pt][\tabcolsep]}l}{\mbox{priority queue\textsuperscript{\citep{pearce_multithreaded_2010, zalewski_distributed_2014}}, dynamic-array buckets\textsuperscript{\citep{madduri_et_al_performance_2014}}}}  \\

    \addlinespace

    \bottomrule
  \end{tabular}
  \caption{Application-Level aspects of DGAs.}
  \label{tab:application-level}
\end{table}
\Cref{tab:application-level} summarizes the set of parameters we identified as the application-level aspects of DGAs, and divide them into 4 categories.  The \emph{approach} category is about the main algorithmic choices, the \emph{algorithmic considerations} category covers the main aspects of the approach, and the categories of \emph{graph representation} and \emph{data structures} cover the data structures that are used.

\subsubsection{Approach}
\label{sec:approach}
The approaches fall into three main categories: \emph{unordered}, \emph{ordered}, and \emph{mixed} (\cf \cref{sec:motivation}). Unordered algorithms expose parallelism and decrease the need for synchronization.  \DC orders work locally, in private priority queues, and HSync does not order work at all (chaotic traversal) in its unordered phase.  Because of the lack of global ordering, both of these approaches are strongly affected by the order of task execution and message order (\eg see \cref{sec:transport}).  \Dstepping, KLA and Hybrid (which executes \Dstepping in unordered phase) utilize rounds of unordered computation separated by global barriers, where the computation is divided into buckets based on a meta property (\eg distance in SSSP) in \Dstepping or on graph topology in KLA.  Thanks to the use of global rounds, these approaches are less susceptible to task and message ordering, but they can exhibit \emph{straggler effect} \citep{zalewski_distributed_2014}  where parts of the system are idle while other parts are still finishing the rounds.  On the other hand, the ordered approaches rely on global barriers to order execution, and because of that they are susceptible to straggler effect. Algorithm load balancing (\cref{sec:algorithmic}) and runtime load balancing (\cref{sec:runtime}) are important characteristics in such algorithms to distribute load evenly among processes and reduce straggler effect.  Ordered algorithms naturally fit the BSP model\citep{valiant_bridging_1990}, but careful implementations can efficiently overlap computation and communication (\eg \citep{checconi_traversing_2014}).

\subsubsection{Algorithmic Considerations}
\label{sec:algorithmic}
\textbf{Data distribution}. There are 3 main ways to distribute graph data among the processing elements. They are \textit{1D distribution},  \textit{2D distribution} and \textit{Edge List Partitioning (ELP)}.

A graph distribution makes use of certain runtime characteristics. For example, 2D distribution based algorithms commonly use collectives from transport (discussed in \Cref{sec:comm-paradigm}). 2D distributions with collectives may suffer from memory scalability issues in small memory machines. To avoid such memory scalability issues, researchers have developed collectives using point-to-point communication~\citep{yoo_scalable_2005}. 

A graph distribution is also connected to the \textit{logical topology} (discussed in \Cref{sec:network-topo}) of the network, which defines the shape of the 2D distribution (square, rectangle etc.).~\citet{buluc_distributed-memory_2015} showed that the logical topology of 2D distribution has an impact on the performance of the algorithm as it effects both communication and computation. They call it \textit{Processor Grid Skewness}. In particular, ``skewness'' changes the number of processors in collective operation and also the size of the local data structure.

~\citet{yoo_scalable_2005} showed that the number of processors participating in collective communication for 1D distribution is $O(p)$ whereas, in 2D distribution, it is reduced to O($\sqrt p$) ($p$ is the number of processors in the process group) and claimed 2D is better for distributed BFS. But recently~\citet{checconi_traversing_2014} compared 2D distribution implemented in~\citep{checconi_breaking_2012} with their 1D distribution and found that the performance of the BFS implementation with 2D distribution is slower compared to 1D because the 2D implementation did not employ runtime characteristics such as compression, communication and computation overlapping methods. Further they also showed scaling was limited in 2D due to cache thrashing.

To overcome certain overhead incurred due to high degree vertices (in 1D),~\citet{pearce_scaling_2013} introduced ELP. For ELP, we sort the edgelist of a graph according to their sources and partition them evenly. But with ELP, a vertex state may reside in multiple nodes (see Delegates under Load Balancing) and requires synchronization of states between nodes via reductions.

\textbf{Optimizations.} Research community in this area uses various techniques to improve the efficiency of a proposed algorithm. 

~\citet{beamer_direction-optimizing_2013} introduced direction optimization to standard level synchronous BFS algorithm. Direction optimization reduces the number of vertices to be visited by traversing vertices in a bottom up fashion.~\citet{beamer_direction-optimizing_2013} showed that direction optimization reduces contention for some atomic operations as bottom up approach removes the need for atomic operations (because only child writes to itself, hence removing the contention). Though direction optimization is a promising approach to improve performance, the initial method suffered from several runtime difficulties~\citep{beamer_distributed_2013}. Each processor needs to check the membership of the frontier set but frontier set is too large to replicate across nodes. In addition, each vertex requires searching for its parent sequentially. To solve the first problem authors of~\citep{beamer_distributed_2013} used 2D distribution and to tackle the second problem they used systolic shifts. 

High degree vertices are challenging for distributed graph algorithms. \textit{Ghost vertices} is an optimization technique to alleviate overhead incurred by high degree vertices . Ghost vertices reduce some remote communication by locally storing the distributed state of high degree vertices. For larger graphs, ghost vertices can limit the memory scalability.

In SSSP, when an algorithm detects a vertex with relatively lower distance the processing node can relax that vertex with high priority. Messages generated using such vertices are called \textit{priority messages}. During the execution of an algorithm, if a node observes that a message with high priority (\eg better vertex distance in SSSP) is received, it is processed immediately bypassing other intermediate data structures. Runtime support is necessary for this optimization to be successful. The priority messages should get delivered to respective owners swiftly. Therefore the sender needs to maintain a separate channel for priority messages. Further if runtime is supporting message aggregation (coalescing), the priority messages should maintain low coalescing buffer size compared to normal buffers.

\textbf{Load Balancing.}
Load balancing in general attempts to distribute work among participating processes uniformly. In the following we discuss few strategies for load balancing.

\textit{Per thread work splitting \& Proxies}:~\citet{chakaravarthy_scalable_2014} used two tier mechanism: intra-node thread level and inter-node vertex-splitting strategies, based on proxies, to achieve load balancing. In intra-node thread level load balancing, besides the owner thread of a high degree vertex, other threads can participate in the relax operation for the edges involved. In-node NUMA characteristics affect intra-node load balancing.  For inter-node processing, the authors employed \textit{proxies}. Proxies reside in different localities than the high degree vertex. Proxies are connected to the high degree vertex with 0 weight. Time for processing a high degree vertex connected with proxies depends on the job placement, i.e. if a proxy is connected through a slow channel, the time to process the vertex relevant to the proxy is increased.

\textit{Random shuffling of vertex identifier}:~\citet{buluc_parallel_2011} achieved a reasonable load balancing by randomly shuffling all the vertex identifiers prior to partition. In the runtime, random shuffling changes the algorithm communication pattern.  This helps to share runtime resources near equally among all processes.

\textit{Delegates}: \citet{pearce_faster_2014} used \textit{delegates} to distribute edge lists of high-degree vertices across  multiple processes. While partitioning, the outgoing edges are placed at the edge's target vertex location. Then one of high degree vertex owners is designated as a \textit{controller} and others are assigned as \textit{delegates}. The controller maintains the state of the vertex and delegates keeps a copy of the updated state. Delegates communicate with each other using asynchronous broadcast and reduction operations. According to the authors of~\citep{pearce_faster_2014}, distributed delegate technique is more efficient compared to ELP in terms of communication reduction. 

\subsubsection{Graph Representation}\label{sec:graph_representation}
In the literature we find two main techniques to represent a graph. They are \textit{Adjacency list} and \textit{Compressed sparse row (CSR)}. The CSR representation minimizes the space needed using a row indexing mechanism. Compared to adjacency list representation, CSR is a more compact and efficient data structure but less versatile.

\citet{edmonds_single-source_2006} showed that, due to compact size (less memory compared to adjacency list) and efficient access methods (direct indexing), the CSR implementation outperforms adjacency list representation. 
Very recently,~\citet{buluc_distributed-memory_2015} showed that, at large scales (33 and above), the CSR representation does not scale with memory. Therefore to represent larger graphs,~\citep{buluc_distributed-memory_2015} used \textit{Doubly Compressed Sparse Column} (DCSC). As per the authors, CSR representation is faster than DCSC. They also demonstrated that CSR performance is affected by in-node multithreading and NUMA behaviors. The authors received 15-17\% performance improvement when executed with multi-threading within NUMA domains with the CSR implementation. 

Storing adjacencies in linked lists incurs cache misses on traversal (pointer chasing of list nodes).  To avoid sequential navigation~\citet{checconi_traversing_2014} used \textit{indexed skip lists} for adjacency lists. Skip lists contain shortcuts to navigate from one source vertex to the following.  When a few vertices need to be visited, a coarse index allows to skip large portions of adjacency list. However Skip Lists need more memory to store additional indexing data (compared to standard adjacency list representation).

\subsubsection{Data Structures for Algorithm Progress}
\label{sec:datastructures}
Algorithms use data structures to maintain intermediate state.  \citet{pearce_scaling_2013} used a \textit{visitor queue} that is implemented as a collection of priority queues, to maintain adjacency vertices of an already visited vertex. A priority queue for a vertex is selected based on a hash function. Using multiple threads with a hash function reduced lock contention.
We used \textit{thread local priority queues} for \DC SSSP~\citep{zalewski_distributed_2014}. We took this approach to avoid contention on priority queues.
To represent frontier vector for BFS,~\citet{buluc_parallel_2011} used a thread-local stack.  These are merged into frontier set at the end of each iteration. Though this approach reduces contention, it needs additional memory to occupy thread-local stacks and copying also takes time. 
In our \Dstepping~\citep{zalewski_distributed_2014} implementation we used a shared memory bucket data structure that is based on arrays. Atomics were used to avoid any race conditions when making changes to buckets.

\subsection{Runtime-Level aspects of DGAs}
\label{sec:runtime}
In~\cref{sec:motivation}, we show how different runtime-level parameters effect the performance of DGAs. In this section, we categorize important runtime-level parameters based on the review of existing literature for DGAs.  ~\Cref{tab:runtime-level} summarizes the set of low-level runtime parameters.


{\providecommand{\mybullet}{\hspace{1em}\raisebox{1pt}{\smaller[4]$\bullet$}\xspace}

\begin{table}
  \centering
  \small
  \setlength{\tabcolsep}{6pt}
  \begin{tabular}{@{}p{.3\linewidth}@{}p{.6\linewidth}@{}}
    \toprule

    \addlinespace
    \mbox{\textbf{Communication Paradigm}} & \multicolumn{1}{@{}r}{\cref{sec:comm-paradigm}}\\
    \midrule
\multicolumn{2}{@{}>{\columncolor{LightGray}[0pt][\tabcolsep]}l}{Point-to-Point\textsuperscript{\citep{checconi_traversing_2014,zalewski_distributed_2014}}, Collectives\textsuperscript{\citep{buluc_distributed-memory_2015, buluc_parallel_2011,checconi_traversing_2014,madduri_et_al_performance_2014}}} \\
    \multicolumn{2}{@{}>{\columncolor{LightGray}[0pt][\tabcolsep]}l}{One-Sided, Active messages\textsuperscript{\citep{zalewski_distributed_2014,chakaravarthy_scalable_2014}}} \\

    \addlinespace[2\defaultaddspace]
    \mbox{\textbf{Transport}} & \multicolumn{1}{@{}r}{\cref{sec:transport}} \\
    \midrule
    \LGlcell{} & \LGrcell{Remotely synchronous}  \\
    \LGlcell{Request Tracking} & \LGrcell{Locally synchronous\textsuperscript{\citep{checconi_traversing_2014}}}  \\
    \LGlcell{} & \LGrcell{Asynchronous\textsuperscript{\citep{zalewski_distributed_2014}}}  \\

    \MGlcell{Progression\textsuperscript{\citep{checconi_traversing_2014,zalewski_distributed_2014}}:} & \MGrcell{}  \\
    \MGlcell{\mybullet Asynchronous} & \MGrcell{System threads\textsuperscript{\citep{zalewski_distributed_2014}}}  \\
    \MGlcell{} & \MGrcell{User threads}  \\
    \MGlcell{\hspace{1em}{\smaller[4]$\bullet$} Synchronous} & \MGrcell{Explicit progress\textsuperscript{\citep{zalewski_distributed_2014}}}  \\ 
    \MGlcell{} & \MGrcell{Runtime scheduler\textsuperscript{\citep{zalewski_distributed_2014}}}  \\ 
    \MGlcell{} & \MGrcell{Lightweight task\textsuperscript{\citep{zalewski_distributed_2014}}} \\ 

    \LGlcell{} & \LGrcell{System Processing Interface (SPI)\textsuperscript{\citep{checconi_breaking_2012,chakaravarthy_scalable_2014,checconi_traversing_2014}}}  \\
    \LGlcell{} & \LGrcell{Message Passing Interface (MPI)\textsuperscript{\citep{buluc_parallel_2011,checconi_traversing_2014,buluc_distributed-memory_2015}}} \\
    \LGlcell{\multirow{-2}{*}{Bit Transport}} & \LGrcell{Active Pebbles (\ampp)\textsuperscript{\citep{willcock_active_2011}}}  \\
    \LGlcell{} & \LGrcell{ARMI\textsuperscript{\citep{harshvardhan_stapl_2013,harshvardhan_kla:_2014}}}  \\

    \MGlcell{Protocol} & \MGrcell{Eager, rendezvous\textsuperscript{\citep{zalewski_distributed_2014}}, completion\textsuperscript{\citep{chakaravarthy_scalable_2014,checconi_traversing_2014}}} \\

    \LGlcell{} & \LGrcell{Message reduction/caching\textsuperscript{\citep{pearce_faster_2014,zalewski_distributed_2014,madduri_et_al_performance_2014}}}  \\
    \LGlcell{Optimization} & \LGrcell{Message coalescing\textsuperscript{\citep{checconi_traversing_2014,pearce_faster_2014,pearce_scaling_2013}}} \\
    \LGlcell{} & \LGrcell{Message compression\textsuperscript{\citep{checconi_traversing_2014}}}  \\

    \MGlcell{Message routing} & \MGrcell{2D, 3D\textsuperscript{\citep{pearce_scaling_2013,pearce_faster_2014}}, ring\textsuperscript{\citep{yoo_scalable_2005}}, hypercube, rook\textsuperscript{\citep{edmonds_expressing_2013}}}  \\

    \LGlcell{Threading} & \LGrcell{Multi-threaded\textsuperscript{\citep{zalewski_distributed_2014}}, serialized, funneled} \\

    \addlinespace[2\defaultaddspace]
    \mbox{\textbf{Network Topology}} & \multicolumn{1}{@{}r}{\cref{sec:network-topo}} \\
    \midrule

    \LGlcell{Physical Topology} & \LGrcell{3D\textsuperscript{\citep{pearce_scaling_2013,checconi_massive_2013}} and 5D\textsuperscript{\citep{checconi_massive_2013}} torus, Dragonfly\textsuperscript{\citep{zalewski_distributed_2014,buluc_distributed-memory_2015}}} \\

    \MGlcell{Logical Topology} & \MGrcell{Skewness\textsuperscript{\citep{buluc_distributed-memory_2015}}, synthetic network\textsuperscript{\citep{pearce_scaling_2013,pearce_faster_2014}}} \\

    \LGlcell{Job topology} & \LGrcell{Job allocation, rank mapping\textsuperscript{\citep{buluc_distributed-memory_2015}}} \\

    \addlinespace[2\defaultaddspace]
    \mbox{\textbf{Local Scheduling}} & \multicolumn{1}{@{}r}{\cref{sec:local-scheduling}} \\
    \midrule

    \LGlcell{Heavyweight} & \LGrcell{Pthreads\textsuperscript{\citep{zalewski_distributed_2014}}, OpenMP\textsuperscript{\citep{buluc_parallel_2011,buluc_distributed-memory_2015}}} \\

    \MGlcell{Lightweight} & \MGrcell{Work stealing, FIFO tasks\textsuperscript{\citep{zalewski_distributed_2014}}} \\

    \LGlcell{Termination} & \LGrcell{Quiescence detection\textsuperscript{\citep{pearce_scaling_2013}}, SKR\textsuperscript{\citep{zalewski_distributed_2014}}} \\

    \MGlcell{Hardware effects} & \MGrcell{L2 atomics\textsuperscript{\citep{chakaravarthy_scalable_2014}}, NUMA effects\textsuperscript{\citep{buluc_distributed-memory_2015}}} \\

    \addlinespace[2\defaultaddspace]
    \mbox{\textbf{Runtime Feedback}} & \multicolumn{1}{@{}r}{\cref{sec:runtime-feedback}} \\
    \midrule

    \multicolumn{2}{@{}>{\columncolor{LightGray}[0pt][\tabcolsep]}l}{\mbox{Optimal K-level\textsuperscript{\citep{harshvardhan_kla:_2014}}, optimal $\Delta$, sync to async switching\textsuperscript{\citep{xie_sync_2015}}}}  \\

    \addlinespace

    \bottomrule
  \end{tabular}
  \caption{Runtime-Level aspects of DGAs.}
  \label{tab:runtime-level}
\end{table}
}

\subsubsection{Communication Paradigm}
\label{sec:comm-paradigm}
The choice of \emph{communication paradigm} can have a notable impact on performance of DGAs.  Each paradigm imposes different tradeoffs in terms of memory constraints, synchronization overhead and network latency.  The \emph{collectives} paradigm is used when large low-overhead stages of all-to-all communication are needed, \emph{point-to-point} paradigm allows for finer overlap between computation and communication at the expense of code complexity, and \emph{active messages} are a refinement of point-point communication that adds an implicit execution of \emph{handlers} on remote objects.  Finally, \emph{one-sided} paradigm provides remote memory operations (GET, PUT, etc.) which are very efficient, but require remote memory management protocol.  For example, collectives are the base of BLAS approaches and level-synchronous approaches.  However, \citet{checconi_traversing_2014} show how using lightweight point-to-point communication may lead to improvements in traditionally synchronous approaches.  They compare their point-to-point implementation using low level SPI interface (discussed in \cref{sec:bit-transport}) to an MPI implementation using collectives.  They note the large memory footprint required for collective buffers, which forces them to decrease the scale of the problem per node.  Furthermore, collectives do not allow for easy interleaving of computation with communication.  Given these two factors and the overhead of MPI over SPI, they note a fivefold decrease in performance.  Active messages are based on point-to-point communication, and they display similar communication performance characteristics (indeed, \cite{checconi_traversing_2014} implements rudimentary active messages using low-level system interfaces).  However, \ampp provides a full scale of active message services, such as message routing, message reduction, and object-based addressing, and automatic execution of handlers~\citep{willcock_am++:_2010}, which requires a runtime scheduler which may have an important impact on the performance of a DGA (\cref{sec:motivation}).

\subsubsection{Transport}
\label{sec:transport}
The transport layer is the part of the stack responsible for sending and receiving bits.  Important properties of transport include how message buffers are handled, which entity manages them, and how frequently they need to be managed. The runtime needs to take several decisions regarding these. 

\textbf{Request Tracking (RT).} refers to how communication requests are made into transport: a request is scheduled and needs to be completed later (\emph{asynchronous}), a request is made and the requester waits until all local data structures can be reused (\emph{locally synchronous}), or a request is made and the requester waits until it has been completely processed (\emph{remotely sychronous}). As an example, \textit{remotely synchronous} request tracking in MPI (\eg by MPI\_Ssend etc.) guarantees a small number of messages on the network, but it hinders parallelization.  On the other hand, using \textit{locally synchronous} may allow more parallelism if the underlying implementation uses an eager protocol (\eg MPI\_Send).  Finally, \textit{asynchronous} RT uses interfaces such as MPI\_Isend/MPI\_IRecv to start requests along with MPI\_Testsome to check for their completion. maximizing overlap between computation and communication. 
However, with the asynchronous RT, the client of a transport must make decisions about how many requests to keep opened at any given time, how many to check, at any given time.  
\citet{checconi_traversing_2014} used asynchronous RT, maintaining one buffer per destination. A send operation queues messages on these buffers until the buffer is full, and then hands it over to the network interface. But before starting to write to the buffer again, their implementation has to wait for the previous send to complete so the buffer becomes available. In~\citep{zalewski_distributed_2014}, \ampp spawns a constant number of receive requests and as many sand requests as necessary, all stored and tracked in one array of MPI requests.

\textbf{Progression.}Completing a round trip through transport requires a protocol for dedicating resources to the transport for bookkeeping, performing bit moving, and delivering the results of completed requests to the application---we call that protocol progression.  Progression influences the timeliness and efficiency of transport delivery, and a wrong progression model can render an application infeasible (\cref{sec:dc-progress}).
In \emph{asynchronous} progression, computing resources are dedicated to make progress.  The resources can be dedicated through \textit{system} or \emph{user} threads.  For example, Cray MPI provides an option for starting progression pthreads that perform internal MPI progress in parallel with the application threads.  \textit{User threads} serve similar purpose but are started by the user explicitly, and, for example, call MPI repeatedly to generate progress.  In contrast, \textit{synchronous} progression is done periodically in the runtime or application level.  In \textit{explicit} asynchronous progress, the application can choose explicitly, bypassing the runtime scheduler (if any), when to call progress, enabling optimizations at the cost of complexity.  For example, we employed explicit polling for our \DC, but we observed observed a decrease in performance.  In a task-based system, network progress can be scheduled as a \emph{lightweight} task.  For example, \ampp implements network polling, buffer flushing, checking for termination, and executing pending handlers for received messages as tasks, on equal footing with application task that run message handlers.  Finally, a runtime with a scheduler, such as \ampp, can perform progress directly in the \emph{scheduler}, which allows for more control and runtime feedback when performing progression.  Most papers do not discuss progression and request tracking explicitly, but the choices made for these parameters may have a profound effect on performance (\cf \cref{sec:dc-progress} for a motivating example).


We illustrate the progression and request tracking by briefly comparing our \DC in \ampp with the work of \citet{checconi_traversing_2014}.
\citeauthor{checconi_traversing_2014} used a lightweight asynchronous communication layer on top of System Processing Interface (SPI), with separate FIFOs for injections and receptions (up to 16 FIFOs each per node, providing network-level parallelism).  They queue messages to per destination buffers where each buffer is exclusively owned and operated on by a single thread, eliminating locking and contention.  Buffers are placed into injection queues when ready, and a thread will wait for completion when another message needs to be sent to a given destination.  \ampp also maintains  per-destination buffers.  The buffers are shared and require atomic operations for writing. Compared to \citeauthor{checconi_traversing_2014}, \ampp does not wait for the send buffers to become available.  Instead, it creates new buffers and can spawn multiple asynchronous send requests for the same destination.  \ampp supports three different progression models.  Polling can be invoked explicitly from the algorithm. Explicit polling bypasses the \ampp task queue, and directly queries the outstanding send and receive requests (with \cplusinl{MPI\_Testsome} on an array of requests).
\ampp also has special purpose user-level tasks that perform progression.  These tasks are executed from \ampp task queue when sends are performed, or when end-of-epoch tests are performed (during these tests \ampp tries to finish an epoch by processing remaining work).

\label{sec:bit-transport}
\textbf{Bit transport.} Bit transport is the lowest-level network interface used by upper levels to deliver bits from one location to another.
In the work from the IBM group \citep{checconi_breaking_2012,chakaravarthy_scalable_2014,checconi_traversing_2014}, the \textit{System Processing Interface (SPI)} communication layer serves as a bit transport (as described above).  The majority of implementations we have discussed use \textit{Message passing Interface} (MPI) for their bit transport.  SPI is a direct interface to hardware queues, while MPI is a complex framework with extra functionality and semantics.  Direct interfaces like SPI may yield more efficient communication, but are less or not at all portable, and may require more implementation effort to implement higher-level features.  The third type of bit transport is based on remote method invocation (RMI) technique and is used in approaches based on \textit{STAPL}~\citep{harshvardhan_stapl_2013,harshvardhan_kla:_2014}, a generic parallel library for graph and other data structures and algorithms. STAPL uses the ARMI (Adaptive Remote Method Invocation) active-message communication library, based on RMI. ARMI supports automatic message coalescing but does not provide routing or message reductions natively.  Both \ampp and ARMI can use different backends, so from the application's perspective they are bit transport themselves, but internally they use a lower-level bit transport.  Such layering makes it hard to optimize parameters related to bit transport (\eg choosing proper coalescing size for \ampp).

\textbf{Protocol.} Bit transport may employ different protocols for different message sizes.  For example, MPI point-to-point communication, may support \emph{eager} protocol for small messages and \textit{rendezvous} protocol for larger transfers, sending messages without or with, respectively, round-trip communication~\citep{mpiprotocoldoc}.  The autonomous and transparent choice of protocols may have a detrimental impact on applications (\eg \cref{section_coalescing_size}).

\textbf{Optimization.} A number of runtime-level optimization techniques have been proposed in the literature to reduce communication overhead and maximize throughput.   In~\cref{sorting_and_caching}, we discussed \textit{message reduction} (caching) in \ampp.  \citet{pearce_faster_2014} used tree based broadcast, reduction and filtering for communication involving high degree vertices. This essentially forces the visitors to traverse the delegate (\cf \cref{sec:algorithmic})  tree and provides the opportunity to filter out messages.  \citet{madduri_et_al_performance_2014} used local lookup arrays to track the tentative distance of every vertex, thus avoiding duplicate request being sent.

Increasing \textit{message coalescing} (\cf \cref{sec:coalescing-size}) size increases the rate at which small messages can be sent over a network at the cost of latency.  \citet{checconi_traversing_2014} used coalescing to pack together all the edges that would be sent to each destination separately and queued them in an intermediate buffer.  \citeauthor{pearce_faster_2014}~\citep{pearce_faster_2014,pearce_scaling_2013} combined coalescing with routing to reduce dense communication. 
Based on the observation that bisection bandwidth becomes a limiting factor for DGAs,~\citet{checconi_traversing_2014} utilized the available processing power to \textit{compress} their buffers of coalesced messages, using differential encoding scheme for compression of vertices.  They reported that compression decreased the number of messages sent in intermediate BFS levels where high message traffic is present, but when the message traffic is low, compression degrades the performance.


\textbf{Message Routing.} ~\citet{pearce_scaling_2013} implemented routing through a synthetic network to mimic the BG/P 3D torus interconnect topology. In a followup paper~\citep{pearce_faster_2014}, the authors additionally embedded the delegate tree as a means for further communication reduction. AM++~\citep{willcock_am++:_2010} also supports two types of software routing strategies: \textit{Rook routing} and \textit{Hypercube routing}. Rook routing reduces the number of communicating buffers to $O(\sqrt p)$~\citep{edmonds_expressing_2013}.  However, a disadvantage of software routing is that it increases message latency.~\citet{yoo_scalable_2005} used ring communication in their optimized collective implementation and adjusted the diameter of the ring to achieve better performance.

\textbf{Threading.} A message passing framework can support different level of thread safety. For MPI, there are 4 levels in total~\citep{mpidoc}: \textit{Single}, \textit{Funneled}, \textit{Serialized} and \textit{Multiple}. In our \DC implementation\citep{zalewski_distributed_2014}, we used Multiple as the threading level with the aim to have the maximum flexibility for multiple threads to invoke MPI functions concurrently. 

\subsubsection{Network Topology}
\label{sec:network-topo}
Computing resources are organized in several specific \textit{physical topologies}: 3D Torus, Dragon Fly, 5D Torus etc. The physical topology has a connection to collectives. For example, MPICH2 provides an all-to-all implementation that is optimized for Aries and Gemini systems. In~\citep{checconi_massive_2013}, the authors stated that the triangle of virtual processors are embedded in the Blue Gene/Q physical topology which is a 5D torus. 

Some of the graph traversal algorithms with 2D distribution make use of the pattern of communication between ranks and that pattern is called \textit{logical topology}. In \cref{sec:algorithmic} we discussed \textit{Processor Grid Skewness}. In relation to skewness, ~\citet{buluc_distributed-memory_2015} found out that the ``tall skinny'' grids performed faster and ``short fat'' grids performed worse than square grids.~\citet{pearce_scaling_2013} used a 3D routing topology to mirror the Blue Gene /P 3D torus interconnect topology. In this way they created a \textit{synthetic network} to implement routing and aggregation.

Job scheduler for computing resources allocate nodes based on scheduling policy and the input request. This formulates the \textit{job topology}.~\citet{Bhatele:2013:GNP:2503210.2503247} showed that performance of an application depends on job placement. Specially in Cray systems the variation can be significant. The variations can be due to the distances among allocated nodes or due to contention on shared network.

\subsubsection{Local Scheduling}
\label{sec:local-scheduling}
Depending on the node-level threading mechanism, thread scheduling policies and synchronization primitives, tasks associated with a DGA can execute in different order with varying frequencies. For example, in an attempt to quickly spread good work, we can send a message with priority and put the message handler infront of the task queue. This is one way to achieve priority scheduling~\citep{zalewski_distributed_2014}. Data structures support (for example bitmaps in sync mode and global queue in async mode in~\citep{xie_sync_2015}) is  also an important factor to achieve effective local scheduling. Below we discuss several thread-granularity and scheduling related factors.

\textit{Heavyweight threads (worker threads)} can be used for intra-node threading.~\citet{buluc_parallel_2011} used MPI for inter node processing and GNU OpenMP for intra-node threading.~\citet{buluc_distributed-memory_2015} also used OpenMP threading in their implementation as it is beneficial for algorithms implemented using BLAS. But \DC based unordered algorithms need more control over threading. Therefore, ~\citep{zalewski_distributed_2014} used a combination of MPI and pthreads.

Lightweight threads, implemented on top of kernel threads, can be scheduled differently. \textit{Lightweight thread scheduling} mechanisms achieve load balancing mostly by \textit{work stealing} and \textit{FIFO} scheduler for tasks. 

The frequency of \textit{termination detection} (TD) is also a very important catalyst for DGA performance. This is especially true for unordered algorithms. In AM++~\citep{zalewski_distributed_2014}, the termination detection is implemented in the runtime level with the help of non-blocking collectives and work balancing FIFO queue. ~\citet{hribar_termination_1998} implemented TD in algorithm level and advocated that, when the computation time becomes smaller than the time it takes to receive a message, high detection frequency should be used. Otherwise low detection frequency should be used.

\textit{Hardware Effects} such as NUMA, L2 atomics impact on the in-node execution of distributed graph algorithms.~\citet{chakaravarthy_scalable_2014} exploited atomic operations implemented in the L2 cache to achieve better aggregate update rate in relaxation.~\citep{buluc_parallel_2011,madduri_experimental_2007,buluc_distributed-memory_2015} also used atomic updates to mitigate synchronization costs. Also,~\citet{buluc_distributed-memory_2015} considered the effect of multithreading within NUMA domains and observed noteworthy performance gain compared to Flat-MPI runs.
 
\subsubsection{Runtime Feedback} 
\label{sec:runtime-feedback}
Choosing algorithmic parameters adaptively, switching between  different algorithms during execution and choosing the mode of algorithms (sync vs async) depend heavily on the feedback provided by the runtime. For example, in~\citep{harshvardhan_kla:_2014}, to adaptively determine and set the level of asynchrony, $k$, in each superstep, a set of conditions are being evaluated to take the decision about whether to double the size of $k$ or not. One of the conditions checks whether the penalty for asynchrony (assessed in terms of wasted work) has exceeded a threshold or not. Based on the support of runtime,  effect/propagation  of wasted work can be mitigated by propagating better work from lower level and thus invalidating wasted work even before processing them in the algorithm level. If this scheme can be put into place in runtime, we can expect different adapted value of $k$ compared to the one without the scheme. Similarly,~\citet{xie_sync_2015} used throughput ie. the amount of vertices processed per unit time as a metric to switch between sync and async mode of algorithms. But the throughput depends on how efficiently the underlying runtime is being employed.
 

\newpage

\section{Conclusions}
\label{sec:recommendations}
We demonstrate clearly and explicitly that the application-level parts that are reported as major contributions do not constitute a complete description of a DGA.  We show how sometimes small changes in runtime can threaten the viability of an approach.  We aim to raise 
awareness of the importance of runtime. With this in mind,  
we first provide a representative overview of application-level aspects in BFS and SSSP.  Then we carefully analyze  runtime aspects, which are usually overlooked. Based on our analysis, we provide the ``anatomy of DGAs''.\MZ{1 dollar bet with martina.}
The anatomy consists of two major layers: the application-level aspects and the runtime-level aspects, which respectively represent the top and the bottom of the software/hardware stack.   Each is further subdivided into categories, and we provide examples from existing research.  We propose a set of guidelines for reporting research design and results.  Altogether, the goal is to make research results in DGAs more accessible, general, and congruent.  To achieve this goal, we believe that research has to be presented in the context of the whole complex stack on modern supercomputers (getting more complex with progress toward exascale).  We intend the guidelines to serve the community as an initial step to iterate and expand on.


Our \cref{tab:application-level,tab:runtime-level} serve as an initial map for \emphbf{reporting the design features}.  It is as important to state which parts of DGAs anatomy are explicitly covered in the results as which are not. Some may remain ``buried in the stack'', their impact unknown (for example, the effects of job placement as in \cref{sec:network-topo} are not usually investigated), and some may not be relevant in a given situation.  Our anatomy helps both consumers and authors of research, the former to understand and the latter to present contributions.

Next, it is important to outline the \emph{\textbf{experimental protocol}} used to obtain the results.  This amounts to stating which parts of the parameter space are covered by experiments and, as importantly, which are not.  Furthermore, it is helpful for the reader to understand \emph{why} certain parts of the parameter space are covered and others are not, even if the reason is as mundane as limited resources (burning through time allocations is easy).  Also, it is often helpful to present negative experimental results if any were observed as they help to uncover to which parameters a given approach is sensitive.

Our analysis and guidelines are intended to be only the first, imperfect step in unifying the field. 
We posit that the DGA research community should collectively develop a set of standards expected from top notch research, acknowledging that DGAs exhibit particularly strong interaction with the software/hardware stack due to their irregularity.  Thus we appeal to the wider community to take our initial suggestion and to help develop standards for more explicit incorporation of runtime interactions in future research results.


\section{Acknowledgments}
This research used resources of NERSC (a DOE Office of Science User Facility Office of Science and U.S. Department of Energy under Contract No. DE-AC02-05CH11231) and Big Red2 (Funded by Lilly Endowment, Inc. and Indiana METACyt Initiative). Research is supported by NSF grant 1111888 and Department of Energy award DE-SC0008809.

\newpage

     \raggedright 
     \sloppy
     \sfcode`\.=1000\relax
\bibliographystyle{abbrvnat}
\bibliography{paper}  

\end{document}

%% file: header.tex
\usepackage[T1]{fontenc}
\usepackage{microtype}
\usepackage[scaled=.84]{beramono}

\usepackage{relsize}
\usepackage{xspace}
\usepackage[normalem]{ulem}

\usepackage[nounderscore]{syntax}

\usepackage[square,numbers,sort&compress]{natbib}
\usepackage{amsmath}
\usepackage{amssymb}
\usepackage{graphicx}
\usepackage{listings}

\usepackage{flushend}

\usepackage{booktabs}
\usepackage{xcolor,colortbl}

\definecolor{LightGray}{gray}{0.97}
\definecolor{MidGray}{gray}{0.9}
\definecolor{HeavyGray}{gray}{0.83}
\newcommand{\lcell}[3]{\multicolumn{1}{@{}>{\columncolor{#1}[0pt][\tabcolsep]}#2}{#3}}
\newcommand{\rcell}[3]{\multicolumn{1}{>{\columncolor{#1}[\tabcolsep][0pt]}#2@{}}{#3}}
\newcommand{\LGlcell}[1]{\lcell{LightGray}{l}{#1}}
\newcommand{\LGrcell}[1]{\rcell{LightGray}{l}{#1}}
\newcommand{\MGlcell}[1]{\lcell{MidGray}{l}{#1}}
\newcommand{\MGrcell}[1]{\rcell{MidGray}{l}{#1}}

\usepackage[
  bookmarksnumbered=true,
  pdfstartview=FitV,
  bookmarksopen,
  bookmarksopenlevel=2]
{hyperref}
\usepackage[all]{hypcap}

\usepackage{wrapfig}

\usepackage{mdwlist}

\lstdefinelanguage{C++1y}{
  alsolanguage=C++,
  escapechar=@,
  breakatwhitespace=true,
  morekeywords = {
    alignof, decltype, concept, axiom, requires, property
  }
}

\lstnewenvironment{program}[1][\small]
{
  \lstset{
    style=C++,
    basicstyle=#1\sffamily,
    keywordstyle=#1\sffamily\bfseries,
    commentstyle=#1\sffamily\itshape,
  }
}
{ }

\lstnewenvironment{programnb}[1][\small]
{
  \lstset{
    style=C++,
    basicstyle=#1\sffamily,
    keywordstyle=#1\sffamily,
    commentstyle=#1\sffamily,
  }
}
{ }

\usepackage{tikz}

\definecolor{lightblue}{rgb}{0,0,0}
\lstdefinelanguage{C++custom}
{
escapeinside={/*@}{@*/},breaklines=true,breakatwhitespace=true,%
basicstyle=\color{lightblue},keywordstyle=,%
lineskip=-.05\baselineskip,morekeywords={pattern,in,is,to,from,out_edges,adj,once,fixed_point,vertex_property,edge_property,vertex,edge,meta,auto,concept,requires,concept_map},%
alsolanguage=C++,
literate = %
  {\{}{\smaller{$\{$}}{1}%
  {\}}{\smaller{$\}$}}{1}%
  {<=}{$\leq$}{1}%
  {-}{--}{1}%
}

\lstdefinestyle{numbers}
{xleftmargin=8pt,numbers=left, numberstyle=\tiny,%
stepnumber=1, numbersep=4pt}

\lstdefinestyle{frametb}
{frame=tb}

\lstdefinestyle{bold-keywords}{keywordstyle=\bfseries}

\lstnewenvironment{myverb}[1][\small]{\lstset{
    columns=fullflexible,
    basicstyle=\color{lightblue}#1\ttfamily,%
    breaklines=true,
  }}{}

\lstnewenvironment{cplus}[1][\footnotesize]{\lstset{language=C++custom,%
    style=numbers,basicstyle=\color{lightblue}#1\ttfamily,%
    keywordstyle=#1\ttfamily,%
    style=bold-keywords,style=frametb}}{}

\lstnewenvironment{cplusuf}[1][\small]{\lstset{language=C++custom,%
    style=numbers,basicstyle=\color{lightblue}#1\ttfamily,%
    keywordstyle=#1\ttfamily,%
    style=bold-keywords,frame=t
  }}{}

\lstnewenvironment{mjava}[1][\small]{\lstset{language=Java,%
    escapeinside={/*@}{@*/},style=numbers,basicstyle=\color{lightblue}#1\ttfamily,%
    keywordstyle=#1\ttfamily,%
    style=bold-keywords,style=frametb}}{}

\lstnewenvironment{cplusnln}[1][\small]{\lstset{language=C++custom,
    xleftmargin=8pt,xrightmargin=8pt,basicstyle=\color{lightblue}#1\ttfamily,%
    keywordstyle=#1\ttfamily,%
    style=bold-keywords,style=frametb}}{}

\makeatletter
\lstnewenvironment{code}[1][\footnotesize]{\lstset{language=C++custom,columns=fullflexible,%
    xleftmargin=8pt,xrightmargin=8pt,basicstyle=\color{lightblue}#1,%
    keywordstyle=#1,style=numbers,%
    style=bold-keywords,mathescape=true}}{\@endparenv} 
\makeatother

\providecommand{\codeinl}[2][\small] 
{{\lstinline[language=C++custom,breaklines=false,columns=fullflexible,%
basicstyle=\color{lightblue}#1\sffamily
,mathescape=true,keywordstyle=#1\sffamily]@#2@}}%

\providecommand{\cplusinl}[2][\normalsize] 
{{\lstinline[language=C++custom,breaklines=false,columns=fullflexible,%
basicstyle=\color{lightblue}#1\ttfamily
,keywordstyle=#1\ttfamily]@#2@}}%

\lstdefinestyle{cppmarkers}{rangeprefix=/*\#\ ,%
includerangemarker=false,%
rangesuffix=\ \#*/}%

\lstdefinelanguage{Haskell-custom}
{
escapeinside={--@}{@--},breaklines=true,breakatwhitespace=true%
language=Haskell,basicstyle=\color{lightblue}\ttfamily,keywordstyle=\ttfamily,%
morekeywords={class,instance,type,newtype,data,where,deriving,import},%
lineskip=-.1\baselineskip,morekeywords={concept,requires,concept_map},
literate={+}{{$+$}}1 {/}{{$/$}}1 {*}{{$*$}}1 {=}{{$=$}}1
               {>}{{$>$}}1 {<}{{$<$}}1 {\\}{{$\lambda$}}1
               {\\\\}{{\char`\\\char`\\}}1
               {->}{{$\rightarrow$}}2 {>=}{{$\geq$}}2 {<-}{{$\leftarrow$}}2
               {=>}{{$\Rightarrow$}}2 
               {\ .}{{$\circ$}}2 {\ .\ }{{$\circ$}}2
               {>>}{{>>}}2 {>>=}{{>>=}}2
               {|}{{$\mid$}}1
             }

\lstnewenvironment{hask}[1][\small]{\lstset{language=Haskell-custom,%
    style=numbers,basicstyle=\color{lightblue}#1\ttfamily,keywordstyle=#1\ttfamily,%
    style=bold-keywords,style=frametb}}{}

\lstdefinestyle{markers}{rangeprefix=\{-\:\ ,%
includerangemarker=false,%
rangesuffix=\ \:-\}}%

\lstdefinestyle{C++}
{
  language=C++1y,
  columns=fullflexible,
  breaklines=true,
}

\newcommand\Dstepping{$\Delta$-stepping\xspace}

\usepackage{mathtools}

\lstdefinelanguage{CASL}{
  escapechar=@,
  breakatwhitespace=true,
  morekeywords = {
    spec, sort, then, op, ops, var, vars, pred, end
  }
}

\lstnewenvironment{casl}[1][\small]
{
  \lstset{
    language=CASL,
    basicstyle=#1\sffamily,
    keywordstyle=#1\sffamily\bfseries,
    commentstyle=#1\sffamily,
    columns=fullflexible,
    breaklines=true,
  }
}
{ }

\usepackage[final]{pdfcomment}
\usepackage[textwidth=1.3\marginparwidth]{todonotes}
\providecommand{\MZ}[1]{\pdfcomment[author=MZ,opacity=.3]{#1}}

\newcommand{\Rplus}{
\raisebox{.3ex}{\smaller
    {\smaller
      \textbf{+}}}}

\newcommand{\ampp}{\mbox{AM\Rplus\Rplus}\xspace}

\newcommand\eg{e.g.,\xspace}

\newcommand\cf{cf.,\xspace}

\usepackage{float}

\floatstyle{ruled}
\newfloat{algorithm}{tbh}{lop}
\floatname{algorithm}{Algorithm}

\usepackage{caption}
\usepackage[subrefformat=parens,labelformat=parens]{subcaption}
\usepackage{multirow}


%% file: reference.tex
\usepackage[sort&compress,capitalize,nameinlink]{cleveref}
\crefname{section}{Sec.}{Secs.}
\Crefname{section}{Sec.}{Secs.}

\providecommand{\todoref}[1]{\hyperref[#1]{todo~\ref*{#1}}}
\providecommand{\mpageref}[1]{\hyperref[#1]{page~\pageref*{#1}}}
\providecommand{\enumref}[1]{\hyperref[#1]{Item~\ref*{#1}}}
\providecommand{\algref}[1]{\hyperref[#1]{Alg.~\ref*{#1}}}
\providecommand{\Algref}[1]{\hyperref[#1]{Algorithm~\ref*{#1}}}

\providecommand{\secref}[1]{\hyperref[#1]{Sec.~\ref*{#1}}}
\providecommand{\Secref}[1]{\hyperref[#1]{\Sec.~\ref*{#1}}}
\providecommand{\secsref}[2]{\hyperref[#1]{Sec.~\ref*{#1}} and \hyperref[#2]{Sec.~\ref*{#2}}}
\providecommand{\Secsref}[2]{\hyperref[#1]{Sec.~\ref*{#1}} and \hyperref[#2]{Sec.~\ref*{#2}}}

\providecommand{\figref}[1]{\hyperref[#1]{Fig.~\ref*{#1}}}
\providecommand{\lstref}[1]{\hyperref[#1]{Listing~\ref*{#1}}}
\providecommand{\lstsref}[2]{\hyperref[#1]{Listings~\ref*{#1}} \hyperref[#2]{and~\ref*{#2}}}
\providecommand{\Lstref}[1]{\hyperref[#1]{Listing~\ref*{#1}}}
\providecommand{\Figref}[1]{\hyperref[#1]{Fig.~\ref*{#1}}}
\providecommand{\figsref}[2]{\hyperref[#1]{Figs.~\ref*{#1}} \hyperref[#2]{and~\ref*{#2}}}
\providecommand{\Figsref}[2]{\hyperref[#1]{Figs.~\ref*{#1}} \hyperref[#2]{and~\ref*{#2}}}
\providecommand{\linesref}[2]{\hyperref[#1]{Lines~\ref*{#1}--\ref*{#2}}}
\providecommand{\linesandref}[2]{\hyperref[#1]{Lines~\ref*{#1}} \hyperref[#2]{and~\ref*{#2}}}
\providecommand{\Linesref}[2]{\hyperref[#1]{Lines~\ref*{#1}} \hyperref[#2]{and~\ref*{#2}}}
\providecommand{\lineref}[1]{\hyperref[#1]{Line~\ref*{#1}}}
\providecommand{\Lineref}[1]{\hyperref[#1]{Line~\ref*{#1}}}
\providecommand{\lemmaref}[1]{\hyperref[#1]{Lemma~\ref*{#1}}}
\providecommand{\Lemmaref}[1]{\hyperref[#1]{Lemma~\ref*{#1}}}
\providecommand{\thmref}[1]{\hyperref[#1]{Theorem~\ref*{#1}}}
\providecommand{\Thmref}[1]{\hyperref[#1]{Theorem~\ref*{#1}}}
\providecommand{\tabref}[1]{\hyperref[#1]{Table~\ref*{#1}}}
\providecommand{\Tabref}[1]{\hyperref[#1]{Table~\ref*{#1}}}
\providecommand{\defref}[1]{\hyperref[#1]{Definition~\ref*{#1}}}
\providecommand{\Defref}[1]{\hyperref[#1]{Definition~\ref*{#1}}}
\providecommand{\exref}[1]{\hyperref[#1]{Example~\ref*{#1}}}
\providecommand{\propref}[1]{\hyperref[#1]{Proposition~\ref*{#1}}}
\providecommand{\appref}[1]{\hyperref[#1]{Appendix~\ref*{#1}}}
\providecommand{\coderef}[1]{\hyperref[#1]{ex.\,{\sffamily\#\,\ref*{#1}}}}
\providecommand{\paperref}[2]{\hyperref[#1]{Paper~\textrm{#2}}}
\providecommand{\papersref}[4]{\hyperref[#1]{Papers~\textrm{#2}} \hyperref[#3]{and~\textrm{#4}}}

\newcommand{\MakeUnaryConcept}[2]{
  \expandafter\newcommand\csname #1\endcsname{\code{#2}\xspace}
  \expandafter\newcommand\csname #1With\endcsname[1]{\code{#2<##1>}}
}

\newcommand{\MakeBinaryConcept}[2]{
  \expandafter\newcommand\csname #1\endcsname{\code{#2}\xspace}
  \expandafter\newcommand\csname #1With\endcsname[2]{\code{#2<##1, ##2>}}
}

\newcommand{\MakeTernaryConcept}[2]{
  \expandafter\newcommand\csname #1\endcsname{\code{#2}\xspace}
  \expandafter\newcommand\csname #1With\endcsname[3]{\code{#2<##1, ##2, ##3>}}
}

\newcommand{\MakeQuaternaryConcept}[2]{
  \expandafter\newcommand\csname #1\endcsname{\code{#2}\xspace}
  \expandafter\newcommand\csname #1With\endcsname[4]{\code{#2<##1, ##2, ##3, ##4>}}
}

\newcommand{\MakeUnaryFunction}[2]{
  \expandafter\newcommand\csname #1\endcsname{\code{#2}\xspace}
  \expandafter\newcommand\csname #1With\endcsname[1]{\code{#2(##1)}}
}

\newcommand{\MakeBinaryFunction}[2]{
  \expandafter\newcommand\csname #1\endcsname{\code{#2}\xspace}
  \expandafter\newcommand\csname #1With\endcsname[2]{\code{#2(##1, ##2)}}
}

\newcommand{\MakeTernaryFunction}[2]{
  \expandafter\newcommand\csname #1\endcsname{\code{#2}\xspace}
  \expandafter\newcommand\csname #1With\endcsname[3]{\code{#2(##1, ##2, ##3)}}
}

\newcommand{\MakeQuaternaryFunction}[2]{
  \expandafter\newcommand\csname #1\endcsname{\code{#2}\xspace}
  \expandafter\newcommand\csname #1With\endcsname[4]{\code{#2(##1, ##2, ##3, ##4)}}
}

\newcommand{\MakeQuinaryFunction}[2]{
  \expandafter\newcommand\csname #1\endcsname{\code{#2}\xspace}
  \expandafter\newcommand\csname #1With\endcsname[5]{\code{#2(##1, ##2, ##3, ##4, ##5)}}
}

\newcommand{\MakeSenaryFunction}[2]{
  \expandafter\newcommand\csname #1\endcsname{\code{#2}\xspace}
  \expandafter\newcommand\csname #1With\endcsname[6]{\code{#2(##1, ##2, ##3, ##4, ##5, ##6)}}
}

\providecommand{\int}{\code{int}\xspace}

\MakeUnaryFunction{isValid}{is\_valid}

\MakeBinaryConcept{Same}{Same}
\MakeBinaryConcept{Derived}{Derived}
\MakeBinaryConcept{Convertible}{Convertible}
\MakeBinaryConcept{Common}{Common}
\MakeBinaryConcept{CommonType}{CommonType}

\MakeUnaryConcept{Integral}{Integral}
\MakeUnaryConcept{SignedIntegral}{SignedIntegral}
\MakeUnaryConcept{UnsignedIntegral}{UnsignedIntegral}

\MakeUnaryConcept{EqualityComparable}{EqualityComparable}
\MakeUnaryConcept{TotallyOrdered}{TotallyOrdered}
\MakeUnaryConcept{Movable}{Movable}
\MakeUnaryConcept{Copyable}{Copyable}
\MakeUnaryConcept{Semiregular}{Semiregular}
\MakeUnaryConcept{Regular}{Regular}

\MakeBinaryConcept{EqualityComparableX}{EqualityComparable}
\MakeBinaryConcept{TotallyOrderedX}{TotallyOrdered}

\MakeBinaryConcept{Function}{Function}
\MakeBinaryConcept{RegularFunction}{RegularFunction}
\MakeBinaryConcept{Predicate}{Predicate}
\MakeBinaryConcept{Relation}{Relation}
\MakeTernaryConcept{RelationX}{Relation}
\MakeBinaryConcept{UnaryOperation}{UnaryOperation}
\MakeBinaryConcept{BinaryOperation}{BinaryOperation}
\MakeTernaryConcept{BinaryOperationX}{BinaryOperation}

\MakeUnaryConcept{Readable}{Readable}
\MakeBinaryConcept{MoveWritable}{MoveWritable}
\MakeBinaryConcept{Writable}{Writable}
\MakeBinaryConcept{IndirectlyMovable}{IndirectlyMovable}
\MakeBinaryConcept{IndirectlyCopyable}{IndirectlyCopyable}

\MakeUnaryConcept{WeaklyIncrementable}{WeaklyIncrementable}
\MakeUnaryConcept{Incrementable}{Incrementable}

\MakeUnaryConcept{WeakInputIterator}{WeakInputIterator}
\MakeUnaryConcept{InputIterator}{InputIterator}
\MakeUnaryConcept{ForwardIterator}{ForwardIterator}
\MakeUnaryConcept{BidirectionalIterator}{BidirectionalIterator}
\MakeUnaryConcept{RandomAccessIterator}{RandomAccessIterator}

\MakeUnaryConcept{ValueType}{ValueType}
\MakeUnaryConcept{DistanceType}{DistanceType}
\MakeUnaryConcept{DifferenceType}{DifferenceType}

\MakeUnaryConcept{Permutable}{Permutable}
\MakeBinaryConcept{PermutableX}{Permutable}
\MakeUnaryConcept{Mergeable}{Mergeable}
\MakeBinaryConcept{MergeableWrt}{Mergeable}
\MakeUnaryConcept{Sortable}{Sortable}
\MakeBinaryConcept{SortableWrt}{Sortable}

\MakeBinaryConcept{RandomNumberGenerator}{RandomNumberGenerator}
\MakeUnaryConcept{UniformRandomNumberGenerator}{UniformRandomNumberGenerator}

\MakeUnaryConcept{AdditiveSemigroup}{AdditiveSemigroup}
\MakeUnaryConcept{AdditiveMonoid}{AdditiveMonoid}
\MakeUnaryConcept{AdditiveCommutativeMonoid}{AdditiveCommutativeMonoid}
\MakeUnaryConcept{AdditiveGroup}{AdditiveGroup}
\MakeUnaryConcept{MultiplicativeSemigroup}{MultiplicativeSemigroup}
\MakeUnaryConcept{MultiplicativeMonoid}{MultiplicativeMonoid}
\MakeUnaryConcept{MultiplicativeGroup}{MultiplicativeGroup}
\MakeUnaryConcept{Semiring}{Semiring}

\MakeUnaryConcept{SemigroupStruct}{Semigroup}
\MakeUnaryConcept{MonoidStruct}{Monoid}
\MakeUnaryConcept{CommutativeMonoidStruct}{CommutativeMonoid}
\MakeUnaryConcept{GroupStruct}{Group}
\MakeUnaryConcept{SemiringStruct}{Semiring}

\MakeBinaryFunction{isWeakRange}{is\_weak\_range}
\MakeBinaryFunction{isCountedRange}{is\_counted\_range}
\MakeBinaryFunction{isBoundedRange}{is\_bounded\_range}
\MakeBinaryFunction{isReadableRange}{is\_readable\_range}
\MakeTernaryFunction{isMovableRange}{is\_movable\_range}
\MakeTernaryFunction{isWritableRange}{is\_writable\_range}
\MakeBinaryFunction{isPermutableRange}{is\_permutable\_range}
\MakeBinaryFunction{isMutableRange}{is\_mutable\_range}

\MakeUnaryFunction{equivalenceRelation}{equivalence\_relation}
\MakeUnaryFunction{strictWeakOrdering}{strict\_weak\_ordering}

\MakeTernaryFunction{notOverlappedForward}{not\_overlapped\_forward}
\MakeTernaryFunction{notOverlappedBackward}{not\_overlapped\_backward}
\MakeTernaryFunction{inRange}{in\_range}
\MakeTernaryFunction{inClosedRange}{in\_closed\_range}

\MakeQuinaryFunction{canMergeRanges}{can\_merge\_ranges}
\MakeSenaryFunction{canMergeRangesWrt}{can\_merge\_ranges}

\MakeBinaryFunction{swap}{swap}

\MakeTernaryFunction{allOf}{all\_of}
\MakeTernaryFunction{anyOf}{any\_of}
\MakeTernaryFunction{noneOf}{none\_of}

\MakeTernaryFunction{find}{find}
\MakeTernaryFunction{findIf}{find\_if}
\MakeTernaryFunction{findIfNot}{find\_if\_not}
\MakeQuaternaryFunction{findFirstOf}{find\_first\_of}
\MakeQuinaryFunction{findFirstOfWrt}{find\_first\_of}

\MakeBinaryFunction{adjacentFind}{adjacent\_find}
\MakeTernaryFunction{adjacentFindWrt}{adjacent\_find}

\MakeTernaryFunction{stdCount}{count}
\MakeTernaryFunction{countIf}{count\_if}

\MakeTernaryFunction{isPermutation}{is\_permutation}
\MakeQuaternaryFunction{isPermutationWrt}{is\_permutation}

\MakeTernaryFunction{isPartitioned}{is\_partitioned}
\MakeTernaryFunction{partitionPoint}{partition\_point}

\MakeTernaryFunction{copyBackward}{copy\_backward}
\MakeTernaryFunction{moveBackward}{move\_backward}

\MakeBinaryFunction{isSorted}{is\_sorted}
\MakeTernaryFunction{isSortedWrt}{is\_sorted}
\MakeBinaryFunction{isSortedUntil}{is\_sorted\_until}
\MakeTernaryFunction{isSortedUntilWrt}{is\_sorted\_until}

\MakeBinaryFunction{iterSwap}{iter\_swap}
\MakeTernaryFunction{swapRanges}{swap\_Ranges}

\MakeTernaryFunction{generate}{generate}

\MakeBinaryFunction{reverse}{reverse}
\MakeTernaryFunction{rotate}{rotate}

\MakeBinaryFunction{sort}{sort}
\MakeTernaryFunction{sortWrt}{sort}
\MakeBinaryFunction{stableSort}{stable\_sort}
\MakeTernaryFunction{stableSortWrt}{stable\_sort}

\MakeTernaryFunction{lowerBound}{lower\_bound}
\MakeTernaryFunction{upperBound}{upper\_bound}
\MakeTernaryFunction{equalRange}{equal\_range}
\MakeTernaryFunction{binarySearch}{binary\_search}

\MakeTernaryFunction{inPlaceMerge}{inplace\_merge}
\MakeQuaternaryFunction{inPlaceMergeWrt}{inplace\_merge}

\MakeBinaryFunction{isHeap}{is\_heap}
\MakeTernaryFunction{isHeapWrt}{is\_heap}
\MakeBinaryFunction{isHeapUntil}{is\_heap\_until}
\MakeBinaryFunction{isHeapUntilWrt}{is\_heap\_until}
\MakeBinaryFunction{pushHeap}{push\_heap}
\MakeTernaryFunction{pushHeapWrt}{push\_heap}
\MakeBinaryFunction{popHeap}{pop\_heap}
\MakeTernaryFunction{popHeapWrt}{pop\_heap}

\MakeBinaryFunction{minElement}{min\_element}
\MakeTernaryFunction{minElementWrt}{min\_element}
\MakeBinaryFunction{maxElement}{max\_element}
\MakeTernaryFunction{maxElementWrt}{max\_element}

\MakeBinaryFunction{nextPermutation}{next\_permutation}
\MakeTernaryFunction{nextPermutationWrt}{next\_permutation}
\MakeBinaryFunction{prevPermutation}{prev\_permutation}
\MakeTernaryFunction{prevPermutationWrt}{prev\_permutation}

\MakeQuaternaryFunction{lexicographicalCompare}{lexicographical\_compare}
\MakeQuinaryFunction{lexicographicalCompareWrt}{lexicographical\_compare}

\MakeTernaryFunction{allEqual}{all\_equal}
\MakeTernaryFunction{anyEqual}{any\_equal}
\MakeTernaryFunction{noneEqual}{none\_equal}

\MakeTernaryFunction{findNth}{find\_nth}
\MakeTernaryFunction{findNthIf}{find\_nth\_if}

\MakeTernaryFunction{countNotEqual}{count\_not\_equal}
\MakeTernaryFunction{countIfNot}{count\_if\_not}

\MakeTernaryFunction{allAdjacent}{all\_equal}
\MakeTernaryFunction{anyAdjacent}{any\_adjacent}
\MakeTernaryFunction{noneAdjacent}{none\_adjacent}

\MakeTernaryFunction{permutations}{permutations}

\MakeTernaryFunction{isHeapOrdered}{is\_heap\_ordered}
\MakeQuaternaryFunction{isHeapOrderedWrt}{is\_heap\_ordered}
